\documentclass[structabstract]{aa}  
\usepackage{graphicx}
\usepackage{epstopdf}
\usepackage{subfigure}
\usepackage{txfonts}
\usepackage[bottom,marginal]{footmisc}
\usepackage{natbib}
\bibpunct{(}{)}{;}{a}{}{,}

\begin{document}
   \title{Interpreting the extended emission around\\ three nearby debris disc host stars}

   \author{Jonathan P. Marshall\inst{1,2,3}
          \and F.~Kirchschlager\inst{4}
          \and S.~Ertel\inst{5}
          \and J.-C.~Augereau\inst{6,7}
          \and G.M.~Kennedy\inst{8}
          \and M.~Booth\inst{9}
          \and S.~Wolf\inst{4}
          \and B.~Montesinos\inst{10}
          \and C.~Eiroa\inst{3}
          \and B.~Matthews\inst{11,12}
          }

   \institute{School of Physics, University of New South Wales, Sydney, NSW 2052, Australia\\
         \email{jonty.marshall@unsw.edu.au}
         \and Australian Centre for Astrobiology, University of New South Wales, Sydney, NSW 2052, Australia
         \and Departamento de F\'isica Te\'orica, Facultad de Ciencias, Universidad Aut\'onoma de Madrid, Cantoblanco, 28049, Madrid, Spain
         \and Christian-Albrechts-Universit\"at zu Kiel, Institut f\"ur Theoretische Physik und Astrophysik, Leibnizstr. 15, 24098 Kiel, Germany
         \and European Southern Observatory, Alonso de Cordova 3107, Vitacura, Casilla 19001, Santiago, Chile 
         \and University Grenoble Alpes, IPAG, F-38000 Grenoble, France
         \and CNRS, IPAG, F-38000 Grenoble, France
         \and Institute of Astronomy, University of Cambridge, Madingley Road, Cambridge, CB3 0HA, UK
         \and Instituto de Astrof\'isica, Pontificia Universidad Cat\'olica de Chile, Vicu\~na Mackenna 4860, 7820436 Macul, Santiago, Chile
         \and Department of Astrophysics, Centre for Astrobiology (CAB,CSIC-INTA), ESAC Campus, P.O. Box 78, 28691 Villanueva de la Ca\~nada, Madrid, Spain
         \and Herzberg Astronomy \& Astrophysics, National Research Council of Canada, 5071 West Saanich Rd, Victoria, BC V9E 2E7, Canada
         \and University of Victoria, Finnerty Road, Victoria, BC, V8W 3P6, Canada
             }

   \date{Received ---; accepted ---}

 
  \abstract
   {Cool debris discs are a relic of the planetesimal formation process around their host star, analogous to the solar system's Edgeworth-Kuiper belt. As such, they can be used as a proxy to probe the origin and formation of planetary systems like our own.}
   {The \textit{Herschel} Open Time Key Programmes `DUst around NEarby Stars' (DUNES) and `Disc Emission via a Bias-free Reconnaissance in the Infrared/Submillimetre' (DEBRIS) observed many nearby, sun-like stars at far-infrared wavelengths seeking to detect and characterize the emission from their circumstellar dust. Excess emission attributable to the presence of dust was identified from around $\sim$~20\% of stars. \textit{Herschel}'s high angular resolution ($\sim$~7\arcsec~FWHM at 100~$\mu$m) provided the capacity for resolving debris belts around nearby stars with radial extents comparable to the solar system (50--100~au).}
   {As part of the DUNES and DEBRIS surveys, we obtained observations of three debris disc stars, HIP~22263 (HD~30495), HIP~62207 (HD~110897), and HIP~72848 (HD~131511), at far-infrared wavelengths with the \textit{Herschel} PACS instrument. Combining these new images and photometry with ancilliary data from the literature, we undertook simultaneous multi-wavelength modelling of the discs' radial profiles and spectral energy distributions using three different methodologies: single annulus, modified black body, and a radiative transfer code.}
   {We present the first far-infrared spatially resolved images of these discs and new single-component debris disc models. We characterize the capacity of the models to reproduce the disc parameters based on marginally resolved emission through analysis of two sets of simulated systems (based on the HIP~22263 and HIP~62207 data) with the noise levels typical of the \textit{Herschel} images. We find that the input parameter values are recovered well at noise levels attained in the observations presented here.}

   \authorrunning{J.P.~Marshall \textit{et al.}} 
   \titlerunning{Extended emission around three nearby disc host stars}
   \maketitle
%

\section{Introduction}

Debris discs around mature, main sequence stars are visible evidence of the occurrence of a planet(esimal) formation process. These discs are most commonly identified through the presence of emission above what is expected from the stellar photosphere at far-infrared wavelengths. The disc is comprised of icy and rocky bodies ranging in size from microns to kilometres, although the observed emission is predominantly produced by dust grains up to a few millimetres in size. Recent and comprehensive reviews of these objects are available in \cite{wyatt08}, \cite{krivov10}, \cite{moromartin13}, and \cite{matthews14}.

The \textit{Herschel} Space Observatory\footnote{\textit{Herschel} is an ESA space observatory with science instruments provided by European-led Principal Investigator consortia and with important participation from NASA.} \citep{herschel_ref}, with its large, 3.5~m mirror and sensitive far-infrared and sub-millimetre instruments the Photodetector Array Camera and Spectrometer \citep[PACS;][]{pacs_ref} and the Spectral and Photometric Imaging REceiver \citep[SPIRE;][]{spire_ref,spire_cal} has provided the capability not only to detect the thermal emission from nearby debris disc stars, but also to measure the spatial extent of debris discs through detection of extended emission in the source radial profiles. Such measurements were only previously possible for the largest and brightest discs thanks to the twin constraints of sensitivity and angular resolution, typically skewing detection of extended emission to younger systems around early type stars. With \textit{Herschel}, the range of debris disc systems has been expanded to observing extended emission from systems with discs with radii $\ge$~30~au and dust fractional luminosities as low as $10^{-6}$ \citep{eiroa13}, demonstrating the capability of \textit{Herschel} to observe discs closely, although not completely, analogous to the solar system's Edgeworth-Kuiper belt \citep{vitense12}. 

The \textit{Herschel} Open Time Key Programmes `DUst around NEarby Stars' \citep[DUNES;][]{eiroa10,eiroa13} and `Disc Emission via a Bias-free Reconnaisance in the Infrared/Submillimetre' \citep[DEBRIS;][]{matthews10} have, in combination, observed a nearly complete volume-limited survey of sun-like stars within 20~pc. The DUNES survey observed 133 stars with the intention of detecting the stellar photospheric emission and measured excess emission from 31, an incidence of 20.2~$\pm$~2.0~\% \citep{eiroa13}. The DEBRIS survey, which observed stars to a uniform depth regardless of target brightness, detected an incidence of 16.5~$\pm$~2.5~\% (Sibthorpe et al., in prep.). The difference in disc detection rate can be attributed to the different survey depths and target distances in the DUNES and DEBRIS surveys. Of the 31 excess stars in the DUNES survey, 16 were measured to be extended compared to the PACS instrument PSF at 100~$\mu$m. 

The three discs analysed here are mature stars previously identified as debris disc hosts through observations by \textit{ISO} \citep[HIP~22263/HD~30495;][]{habing01} and \textit{Spitzer} \citep[HIP~62207/HD~110897, HIP~72848/HD~113511;][]{trilling08}. The additional information on the dust spatial location provided by the new PACS images, in combination with extending or more densely sampling the source spectral energy distributions (SEDs) at far-infrared wavelengths, has presented the opportunity to greatly improve our understanding of the constituent dust in these debris discs through simultaneous modelling of both the disc thermal emission and radial brightness profiles. Analysis of resolved debris discs by such methods has been extremely successful in constraining the properties of dust grains around a broad range of stars e.g. HD~207129 \citep{loehne12}, HD~181327 \citep{lebreton12}, 49 Ceti \citep{roberge13}, HD~32297 \citep{donaldson13}, HIP~17439 \citep{ertel14}, and HD~109085 \citep{duchene14}, shedding new light on the dust dynamics, grain composition and revealing the important physical processes in these systems. 

In Section \ref{sect_obs} the \textit{Herschel} observations of the three targets along with the stellar parameters and photometry taken from the literature used to fit each target are presented. In Section \ref{sect_mod}, the three modelling approaches are summarized. In Section \ref{sect_res}, the results are presented along with a discussion of these new modelling results and their impact on the interpretation of these debris disc stars. In Section \ref{sect_int} we assess the impact that accurate determination of the disc extent has on recovery of the disc structure and dust parameters. Finally, in Section \ref{sect_con}, we summarize our findings and present our conclusions.

\section{Observations and analysis} \label{sect_obs}
\begin{figure*}[ht!!]
\centering
\includegraphics[width=0.8\textwidth,clip,trim = 0mm 0mm 15mm 0mm]{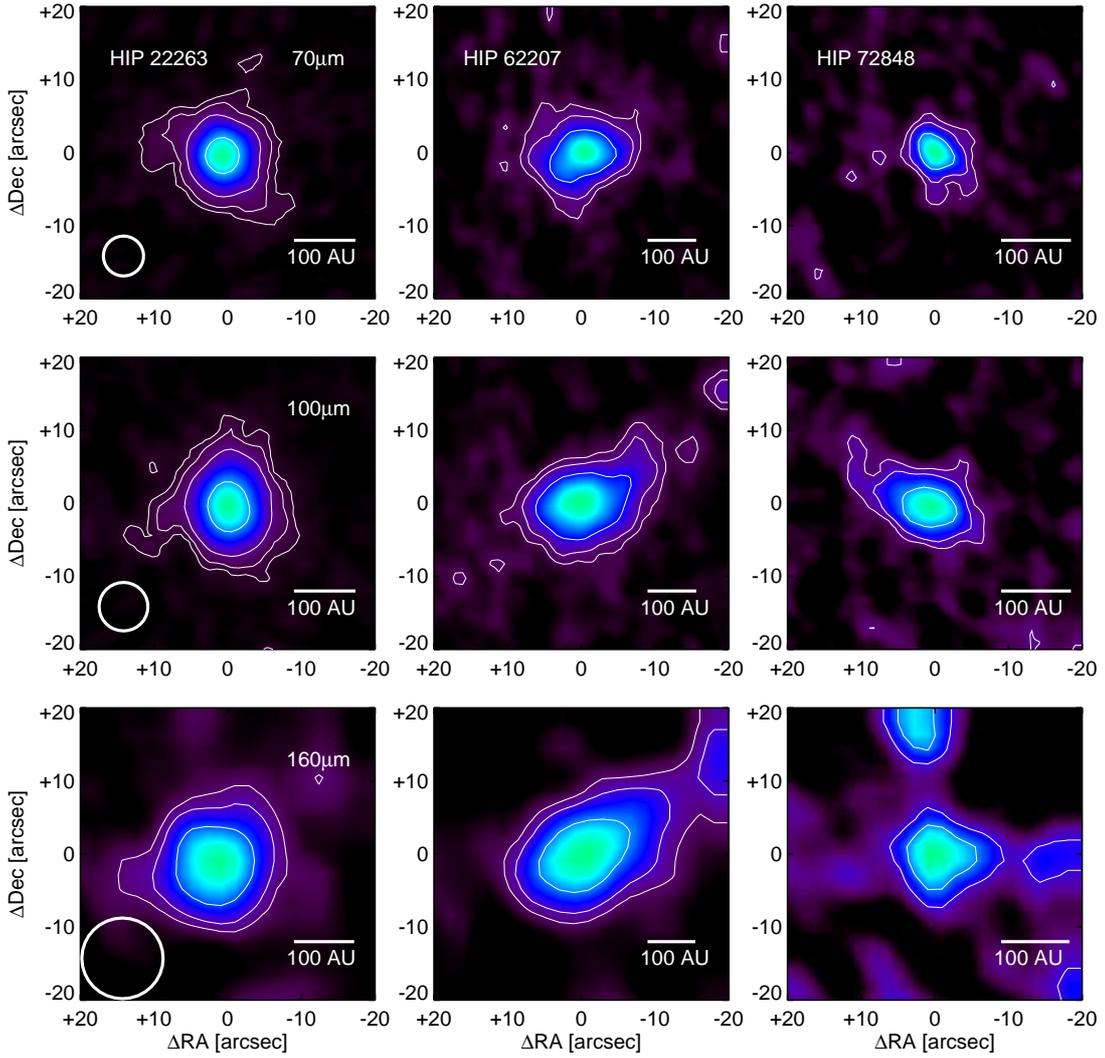}
\caption{\textit{Herschel} PACS images of HIP~22263, HIP~62207 and HIP~72848 (left to right) at 70, 100 and 160~$\mu$m (top to bottom). Contours are at increments of 3-, 5-, 10- and 30-$\sigma$ from the image background level. Colour scale is linear from the background to source peak brightness. Orientation is north up, east left. Pixel scales are 1\arcsec~per pixel at 70 and 100~$\mu$m, and 2\arcsec~per pixel at 160~$\mu$m. The PACS beam size in each band is represented by the circle in the bottom left corner of the images in the left hand column.}
\label{disc_imgs}
\end{figure*}

PACS scan map observations of the targets were taken with both the 70/160 and 100/160 channel combinations. The observations were carried out following the recommended parameters laid out in the scan map release note, i.e. each scan map consisted of 10 legs of 3$'$ length, with a 4\arcsec~separation between legs, scanning at the medium slew speed (20\arcsec~per second). Each target was observed at two array orientation angles (70\degr~and 110\degr) to improve noise suppression and assist in the removal of low frequency (1/$f$) noise, instrumental artefacts and glitches during image reconstruction. A SPIRE small map observation of HIP~22263 was taken covering a region 4$'$ around the target with an equivalent on-source time of 134~s. Neither HIP~62207 nor HIP~72848 were observed with SPIRE due to the expected faintness of their respective discs at sub-mm wavelengths. A summary of the \textit{Herschel} observations can be found in Table \ref{obs_log}. 

\begin{table}[ht]
\centering
\caption{Observation log.}
\begin{tabular}{lllrrr}
\hline\hline
\multicolumn{1}{c}{HIP} & \multicolumn{1}{c}{HD} &  \multicolumn{1}{c}{Instrument$^{a}$} & \multicolumn{1}{c}{Obs. ID} & \multicolumn{1}{c}{OD$^{b}$} & \multicolumn{1}{c}{OT$^{c}$} \\
   &    &  \multicolumn{1}{c}{and Bands [$\mu$m]}  & \multicolumn{1}{c}{[1342$\ldots$]} &          & \multicolumn{1}{c}{[s]} \\
\hline
22263 & 30495 & P~\phantom{0}70/160 & 238836/7 & 1001 & 288 \\                             
\ldots & \ldots & P~100/160 & 193112/3 & \phantom{0}321 & 1440 \\                             
\ldots & \ldots & S~250/350/500 & 203629 & \phantom{0}467 & 134 \\
62207 & 110897 & P~\phantom{0}70/160 & 237972/3 & \phantom{0}967 & 288 \\
\ldots & \ldots & P~100/160 & 212391/2 & \phantom{0}604 & 1440 \\
72848 & 131511 & P~\phantom{0}70/160 & 237142/3 & \phantom{0}973 & 288 \\
\ldots & \ldots & P~100/160 & 212768/9 & \phantom{0}613 & 1440 \\
\hline
\end{tabular}
\tablefoot{(a) P = PACS, S = SPIRE; (b) Operation Day; (c) On-source integration time.}
\label{obs_log}
\end{table}

\subsection{Images}

All data reduction was carried out in the \textit{Herschel} Interactive Processing Environment, user release version 10.0.0 \citep[HIPE,][]{ott10}. The PACS observations were reduced using calibration version 45. Individual PACS scans were processed with a high pass filter to remove background structure, using high pass filter radii of 15 frames at 70~$\mu$m, 20 frames at 100~$\mu$m, and 25 frames at 160~$\mu$m, suppressing structure larger than 62\arcsec, 82\arcsec, and 102\arcsec~in the final images, respectively. To avoid removing source flux, a region 20\arcsec~in radius centred on the source peak was masked from the filtering process. Deglitching was carried out using the second level spatial deglitching task with a threshold of 10-$\sigma$. Final image scales were 1\arcsec~per pixel at 70 and 100~$\mu$m, and 2\arcsec~per pixel at 160~$\mu$m, compared to native instrument pixel sizes of 3.2 (70 and 100~$\mu$m) and 6.4\arcsec (160~$\mu$m). \textit{Herschel} PACS images of the targets are presented in Fig. \ref{disc_imgs}.

The SPIRE observation of HIP~22263 was also reduced in HIPE, using SPIRE calibration version 8.1; the small map was created using the standard pipeline routine, using the naive mapper option. Image scales of 6\arcsec, 10\arcsec, and 14\arcsec~per pixel were used at 250, 350, and 500~$\mu$m.

\subsection{Stellar parameters}

The stellar parameters used in the fitting process are summarized in Table \ref{star_params}. Distances were taken from the re-reduction of the HIPPARCOS data by \cite{vanleeuwen07}. The adopted effective temperature, $T_{\rm eff}$, surface gravity, log$~g$, and metallicity, [Fe/H], values are averages of the photometric and spectroscopic values taken from \cite{gray03}, \cite{santos04}, \cite{takeda05}, \cite{vf05}, \cite{gray06}, \cite{fuhrmann08}, \cite{sousa08}, and \cite{holmberg09}. Rotational periods, $P_{\rm rot}$, are taken from \cite{baliunas83,baliunas96}, and \cite{gaidos00} (HIP~22263), \cite{sf87} (HIP~62207), and \cite{henry95} (HIP~72848). Activity indices, $\log R^{'}_{\rm HK}$, are taken from \cite{martinezarnaiz10}. Bolometric luminosities and stellar radii were estimated from the absolute magnitude and bolometric corrections using measurements by \cite{flower96}, whilst X-ray luminosities were calculated from a combination of \textit{ROSAT}, \textit{XMM} and \textit{Chandra} data. Ages were calculated using the available $\log R^{'}_{\rm HK}$ activity indicies, X-ray luminosities and rotational period data using the same approach as in \cite{maldonado10}. A large scatter between individual age measurements is typical as these stars are on the main sequence and therefore age determination is strongly susceptible to effects of $T_{\rm eff}$ and [Fe/H].

The stellar photosphere contribution to the total flux density was computed, using the stellar parameters, from a synthetic stellar atmosphere model interpolated from the PHOENIX/\textit{Gaia} grid \citep{brott05}. Optical and near infrared photometry including Stromgren $uvby$ \citep{HauckMerm1998}, {\sc Hipparcos} $BVI$ \citep{Perryman1997}, and 2MASS $JHK_{s}$ \citep{Skrutskie2006} measurements constrain the stellar component of the SED, which has been scaled to the optical, near infrared and unsaturated \textit{WISE} \citep{wright10} photometry, following the method of \cite{bertone04}. Further details of the photosphere fitting process can be found in Appendix C of \cite{eiroa13}.

\begin{table}[ht]
\centering
\caption{Stellar physical properties. Coordinates are ICRS J2000.}
\tabcolsep 2 pt
\begin{tabular}{lrrr}
\hline\hline
Parameter & HIP~22263 & HIP~62207 & HIP~72848 \\
\hline
Distance [pc]  & 13.28~$\pm$~0.07 & 17.38~$\pm$~0.10 & 11.51~$\pm$~0.06 \\
Right Ascension [h:m:s] & \phantom{ }04 47 36.29 & \phantom{ }12 44 59.41 & \phantom{ }14 53 23.77  \\
Declination [d:m:s] & -16 56 04.04 & +39 16 44.11 & +19 09 10.08 \\
Spectral type & G3~V & G0~V & K2~V \\
$V$, $B-V$ [mag] & 5.49, 0.63  & 5.95, 0.56 & 6.00, 0.84 \\
$M_V$, B.C.$^{a}$ [mag] & 4.87, -0.08  & 4.76, -0.04 & 5.70, -0.23 \\
$L_{\star}$ [L$_{\odot}$] & 0.951 & 1.055 & 0.498 \\ 
$T_{\rm eff}$ [K] & 5814 & 5860 & 5313 \\
$\log g$ & 4.47  & 4.33 & 4.57 \\
$R_{\star}$ [$R_{\odot}$] & 1.00 & 1.00 & 0.86 \\
$M_{\star}$ [$M_{\odot}$] & 1.00 & 0.86 & 0.84 \\
$[\rm{Fe/H}]$ & 0.01 & -0.53 & 0.10 \\
$\upsilon \sin i$ [kms$^{-1}$] & 2.9 & 3.0 & 4.0 \\
$P_{\rm rot}$ [days] & 7.6 & 13.0 & 10.4 \\
$\log R'_{\rm{HK}}$ & -4.51 & -4.98 & -4.52 \\
$\log~L_{\rm{X}}/L_{\star}$ & -4.73 & \ldots & -4.8 \\
$P_{\rm{rot}}$ age [Gyr] & 1.12 & 2.19 & 0.57 \\
$R'_{\rm{HK}}$ age [Gyr] & 1.21 & 6.24 & 0.69 \\
$L_{\rm{X}}/L_{\star}$ age [Gyr] & 0.44 & \ldots & 0.70 \\
\hline
\end{tabular}
\tablefoot{(a) Bolometric correction.}
\label{star_params}
\end{table}

\subsection{Disc photometry}

For the purposes of SED modelling, the PACS and SPIRE photometry were supplemented by a range of infrared observations including mid-infrared photometry from the \textit{AKARI} IRC \citep[9/18~$\mu$m][]{irc_psc} and \textit{WISE} \citep[][]{wright10} all sky surveys, \textit{Spitzer} IRS \citep{houck04} spectra for HIP~22263 and HIP~62207, \textit{Spitzer} MIPS \citep{rieke04} 24 and 70~$\mu$m flux densities \citep{eiroa13} and SCUBA sub-mm photometry \citep[for HIP~22263, ][]{greaves09}. The \textit{Herschel} photometry and all complementary data are summarized in Table \ref{ir_phot}, whilst the best-fit model SEDs are presented in Fig. \ref{grater_plots}.

\subsubsection{Spitzer}

The MIPS 24 and 70 photometry were taken from the DUNES archive. The original data (HIP~22263 and HIP~62207, program ID: 41, PI: Rieke, G.; HIP~72848, program ID: 30490, PI: Koerner D.~W.) having been reduced similarly to the method outlined in \cite{bryden09} and the flux densities measured by aperture photometry using aperture radii of 15\farcs3 and 14\farcs8, background annuli of 30\farcs6--43\farcs4 and 39\farcs4--78\farcs8 and aperture corrections of 1.15 and 1.79 at 24 and 70~$\mu$m, respectively. None of the three discs exhibit significant warm excess emission at wavelengths $<~25~\mu$m. A colour correction factor of 0.893 was assumed for the dust component of the measured flux densities at 70~$\mu$m, assuming the source temperature to be $\sim$~50~K (colour correction for the stellar component is unity).

The IRS spectrum of HIP~22263 (program ID: 50150, PI: Rieke, G.) was taken from CASSIS\footnote{The Cornell Atlas of \textit{Spitzer}/IRS Sources (CASSIS) is a product of the Infrared Science Center at Cornell University, supported by NASA and JPL.} \citep{lebouteiller11} spanning 13--39~$\mu$m, whilst that of HIP~62207 (program ID: 20463, PI: Ciardi, D.) was taken from the DUNES archive\footnote{http://sdc.cab.inta-csic.es/dunes/jsp/masterTableForm.jsp} \citep{eiroa13}, spanning 8--39~$\mu$m. The reduced spectra were scaled by matching the IRS spectrum to the predicted photosphere model at wavelengths shorter than 20~$\mu$m through a least-squares fit, requiring application of scaling factors of $<$~10\% in both cases. Flux densities were estimated from the spectra at 24, 32 and 37~$\mu$m by averaging of the spectrum within bounds of $\pm$~2~$\mu$m around the given wavelength.

\subsubsection{Herschel}

For PACS, the flux densities were measured using circular apertures with radii of 15\arcsec~at 70 and 100~$\mu$m, and 20\arcsec~at 160~$\mu$m. The sky background and r.m.s. scatter was estimated from the mean and standard deviation of the total flux density in twenty five square apertures with the same area as the flux aperture randomly spaced around the source position at separations of 30--60 pixels. Values were corrected for the both the aperture size (based on the encircled energy fraction measured for a point source) and the source colour, assuming a 5000~K black body for the stellar contribution (factors of 1.016, 1.033, and 1.074 at 70--160~$\mu$m) and a 50~K black body for the disc contribution (factors of 0.982, 0.985, and 1.010 at 70--160~$\mu$m). 

For the SPIRE observation, the source flux density and sky background were measured using the sextractor tool in HIPE. These flux densities were colour corrected using appropriate factors given in the SPIRE photometry release note assuming the slope of the source was the Rayleigh-Jeans tail of a black body (factors of 0.9417, 0.9498, and 0.9395 at 250--500~$\mu$m).

\begin{table*}[ht]
\centering
\caption{Photometry used in source modelling.}
\begin{tabular}{cccclc}
\hline\hline
Wavelength & HIP~22263 & HIP~62207 & HIP~72848 & Instrument & Reference\\
$[\mu \rm{m}]$  & \multicolumn{3}{c}{Flux Density [mJy]} & & \\       
\hline
\phantom{00}0.349 & \phantom{0}5077~$\pm$~\phantom{0}433 & \phantom{0}4150~$\pm$~\phantom{0}353 & \phantom{0}1842~$\pm$~\phantom{0}158 & Str\"omgren \textit{u} & 1 \\
\phantom{00}0.411 & 12270~$\pm$~\phantom{0}510 & \phantom{0}8925~$\pm$~\phantom{0}370 & \phantom{0}5508~$\pm$~\phantom{0}230 & Str\"omgren \textit{v} & 1 \\
\phantom{00}0.440 & 15180~$\pm$~\phantom{0}280 & 10600~$\pm$~\phantom{0}195 & \phantom{0}7824~$\pm$~\phantom{0}144 & Johnson \textit{B} & 2 \\
\phantom{00}0.466 & 18850~$\pm$~\phantom{0}349 & 12610~$\pm$~\phantom{0}234 & 10710~$\pm$~\phantom{0}198 & Str\"omgren \textit{b} & 1 \\
\phantom{00}0.546 & 23580~$\pm$~\phantom{0}435 & 15440~$\pm$~\phantom{0}284 & 14740~$\pm$~\phantom{0}272 & Str\"omgren \textit{y} & 1 \\
\phantom{00}0.550 & 23180~$\pm$~\phantom{0}427 & 15170~$\pm$~\phantom{0}280 & 14490~$\pm$~\phantom{0}267 & Johnson \textit{V} & 2 \\
\phantom{00}0.790 & 30660~$\pm$~\phantom{0}565 & 19340~$\pm$~\phantom{0}356 & 23690~$\pm$~\phantom{0}436 & Cousins \textit{I} & 2 \\
\phantom{00}1.235 & 26070~$\pm$~6154 & 13590~$\pm$~2340 & 17770~$\pm$~4296 & 2MASS \textit{J} & 3 \\
\phantom{00}1.662 & 23120~$\pm$~5064 & 13920~$\pm$~4615 & 22100~$\pm$~7328 & 2MASS \textit{H} & 3 \\
\phantom{00}2.159 & 16760~$\pm$~\phantom{0}556  & 10910~$\pm$~2916 & 12520~$\pm$~3816 & 2MASS \textit{K}$_{\rm s}$ & 3 \\
\phantom{00}3.353  & \phantom{0}8020~$\pm$~\phantom{0}762 & \phantom{0}5160~$\pm$~\phantom{0}418 & \phantom{0}8700~$\pm$~\phantom{0}899 & \textit{WISE} W1 & 4 \\
\phantom{00}9\phantom{.000}  & \phantom{0}1460~$\pm$~\phantom{00}87 & \phantom{00}917~$\pm$~\phantom{00}58 & \phantom{0}1530~$\pm$~\phantom{00}91 & \textit{AKARI} IRC9 & 5 \\
\phantom{0}11.561 & \phantom{00}742~$\pm$~\phantom{00}10 & \phantom{00}464~$\pm$~\phantom{000}6 & \phantom{00}796~$\pm$~\phantom{00}10 & \textit{WISE} W3 & 4 \\ 
\phantom{0}12\phantom{.000} & \phantom{0}1120~$\pm$~\phantom{00}56 & \phantom{00}659~$\pm$~\phantom{00}53 & \phantom{0}1130~$\pm$~\phantom{0}124 & \textit{IRAS} 12 & 6 \\
\phantom{0}18\phantom{.000} & \phantom{00}310~$\pm$~\phantom{00}20 & \phantom{00}178~$\pm$~\phantom{00}23 & \phantom{00}384~$\pm$~\phantom{00}26 & \textit{AKARI} IRC18 & 5 \\ 
\phantom{0}22.088 & \phantom{00}226~$\pm$~\phantom{000}4 & \phantom{00}138~$\pm$~\phantom{000}3 & \phantom{00}239~$\pm$~\phantom{000}4 & \textit{WISE} W4 & 4 \\
\phantom{0}24\phantom{.000} & \phantom{00}186~$\pm$~\phantom{000}4 & \phantom{00}111~$\pm$~\phantom{000}2 & \phantom{00}193~$\pm$~\phantom{000}4 & \textit{Spitzer} MIPS24 & 7 \\
\phantom{0}25\phantom{.000} & \phantom{00}224~$\pm$~\phantom{00}27 & \phantom{00}159~$\pm$~\phantom{00}45 & \phantom{00}281~$\pm$~\phantom{00}48 & \textit{IRAS} 25 & 6 \\
\phantom{0}32\phantom{.000} & \phantom{00}120~$\pm$~\phantom{00}24 & \phantom{000}65~$\pm$~\phantom{000}3 & \ldots & \textit{Spitzer} IRS & 8 \\
\phantom{0}37\phantom{.000} & \phantom{00}115~$\pm$~\phantom{0}125 & \phantom{000}52~$\pm$~\phantom{000}9 & \ldots & \textit{Spitzer} IRS & 8 \\
\phantom{0}70\phantom{.000} & \phantom{00}127.6~$\pm$~8.9 & \phantom{00}62.4~$\pm$~\phantom{0}6.5 & \phantom{00}42.7~$\pm$~\phantom{0}6.7 & \textit{Spitzer} MIPS70 & 7 \\
\phantom{0}70\phantom{.000} & \phantom{00}128.6~$\pm$~6.7 & \phantom{00}76.5~$\pm$~\phantom{0}6.1 & \phantom{00}36.8~$\pm$~\phantom{0}3.8 & \textit{Herschel} PACS70 & 8 \\
100\phantom{.000}  & \phantom{00}78.3~$\pm$~\phantom{0}4.4 & \phantom{00}55.6~$\pm$~\phantom{0}6.6 & \phantom{00}25.0~$\pm$~\phantom{0}2.1 & \textit{Herschel} PACS100 & 8 \\
160\phantom{.000}  & \phantom{00}46.3~$\pm$~\phantom{0}3.8 & \phantom{00}43.9~$\pm$~\phantom{0}3.2 & \phantom{00}14.4~$\pm$~\phantom{0}3.2 & \textit{Herschel} PACS160 & 8 \\
250\phantom{.000}  & \phantom{00}21.8~$\pm$~\phantom{0}6.8 & \ldots & \ldots & \textit{Herschel} SPIRE250 & 8 \\
350\phantom{.000}  & \phantom{00}16.4~$\pm$~\phantom{0}6.9 & \ldots & \ldots & \textit{Herschel} SPIRE350 & 8 \\
500\phantom{.000}  & $<$~24.7\phantom{00000} & \ldots & \ldots & \textit{Herschel} SPIRE500 & 8 \\
850\phantom{.000}  & \phantom{000}5.6~$\pm$~\phantom{00}1.7 & \ldots & \ldots & JCMT SCUBA850 & 9 \\
\hline
\end{tabular}
\tablebib{1. \cite{HauckMerm1998}; 2. \cite{Perryman1997}; 3. \cite{Skrutskie2006} ; 4. \cite{wright10}; 5. \cite{irc_psc}; 6. \cite{iras_psc}; 7. \cite{eiroa13}; 8. This work; 9.\cite{greaves09}.}
\label{ir_phot}
\end{table*}

\subsection{Radial profiles}

As part of the disc modelling process, radial profiles of the disc were taken along the semi-major and semi-minor axis in all three PACS bands. The radial profiles were measured in the following way: firstly, the image of each disc was rotated according to the position angle from the fitted ellipse such that the semi-major axis lay along the image \textit{x}-axis. The rotated image was then interpolated onto a grid ten times finer than the original image, i.e. 0.1\arcsec~(at 70 and 100~$\mu$m) or 0.2\arcsec~(at 160~$\mu$m). The radial profile was measured by taking the mean at distances equivalent to 1\arcsec~(at 70 and 100~$\mu$m) or 2\arcsec~(at 160~$\mu$m) from the disc centre along the major and minor axes in both positive and negative directions using squares of 11$\times$11 grid points. The uncertainty in this mean value was measured by combining in quadrature the difference in the values at positive and negative offset from the disc centre and sky background measured through aperture photometry. The PSF model, a scaled and rotated image of $\alpha$ Bo\"otis, was treated in the exact same manner and the PSF radial profiles were compared to those of the disc to see if it was significantly extended. The PSF measurements for each target were made in the same angle relative to the telescope (the telescope roll angle varies between ODs) to mitigate the effect of the non-circular PACS beam on the measured profiles. It is noted that the reproducibility of the PACS PSF is uncertain at the 10~\% level for 70~$\mu$m images and 4~\% at 100~$\mu$m \citep{kennedy12b}. Since our targets were all extended in both of the shorter wavelength PACS bands, the extended emission we measure is robust to the choice of PSF model.

The radial profiles are presented in Fig. \ref{grater_plots} and a summary of the observational properties of the discs in all three PACS bands is presented in Table \ref{orient}. All three of the targets exhibit extended emission as a broadened PSF rather than a ring-like annular structure. The angular resolution is too low and uncertainties too large to identify any flattening, or dip, in the source brightness profiles close to their peak which might indicate the presence of an inner hole to the discs examined here. We fit a rotated 2D Gaussian profile to the source, weighted by the observation's error map, in order to estimate the source extent along the major and minor axes, $A_{\rm image}$ and $B_{\rm image}$, respectively, along with the disc inclination, $i$, and position angle, $\theta$. Due to the low S/N and marginal extension of the discs presented here, the real source size is estimated solely through subtraction from the PACS beam rather than by image processing through one or more deconvolution methods, as has been undertaken for several of the better resolved, brighter DUNES targets, e.g. HIP~7978 \citep{liseau10}, HD~207129 \citep{marshall11,loehne12}, or HIP~17439 \citep{ertel14}. 

For HIP~22263, we recover extended source brightness profiles along the disc major and minor axes at both 70 and 100~$\mu$m, but the disc is only extended along the major axis at 160~$\mu$m. For HIP~62207, the disc is extended along its major axis in all three PACS bands, and along the minor axis in the two shorter wavelength images. For HIP~72848, the disc is extended along the major axis at 70~$\mu$m and 100~$\mu$m, but point-like at 160~$\mu$m and is not extended along the minor axis in any of the three images. We interpret the lack of extended emission from HIP~22263 and HIP~72848 at 160~$\mu$m to be due to the disc size being comparable to the PACS beam FWHM at that wavelength, although this assessment is complicated in the case of HIP~72848 by the degree of background structure at 160~$\mu$m. The inclination of each disc is estimated from the ratio of the semi-minor to semi-major axes, i.e. $\cos i\!=\!B_{\rm image}/A_{\rm image}$, after deconvolution of the observed image from the instrument PSF represented by an appropriate observation of $\alpha$ Bo\"otis reduced in the same fashion as the observations and rotated to the same orientation as the observed images. Our resulting estimates of the disc inclinations are consistent with those of \cite{greaves14}, which also used the \textit{Herschel} PACS images of each target to derive the disc orientation. The disc position angle is measured from the rotation of the 2D Gaussian east of north. 

\begin{table}
\centering
\caption{Disc orientation and extent.}
\label{orient}
\begin{tabular}{lccc}
\hline\hline
          & \multicolumn{3}{c}{HIP~22263}\\
Parameter & 70~$\mu$m & 100~$\mu$m & 160~$\mu$m \\
\hline
A$_{\rm image}$ [\arcsec] & \phantom{0}8.5~$\pm$~0.1 & 10.2~$\pm$~0.1 & 13.6~$\pm$~0.3 \\
B$_{\rm image}$ [\arcsec] & \phantom{0}7.6~$\pm$~0.1 & \phantom{0}8.2~$\pm$~0.1 & 11.1~$\pm$~0.3 \\
PSF [\arcsec] & 5.6 & 6.8 & 11.3 \\
A$_{\rm disc}$ [\arcsec] & \phantom{0}5.8~$\pm$~0.6 & \phantom{0}6.1~$\pm$~0.3 & \phantom{0}4.8~$\pm$~0.5 \\
A$_{\rm disc}$ [au] & \phantom{0}77.1~$\pm$~8.0 & \phantom{0}81.3~$\pm$~4.0 & \phantom{0}63.8~$\pm$~6.7 \\
$\theta$ [\degr] & \phantom{0}28~$\pm$~5 & \phantom{00}5~$\pm$~2 & \phantom{0}29~$\pm$~9 \\
$i$ [\degr] &\multicolumn{3}{c}{51~$\pm$~10}\\
\hline
          & \multicolumn{3}{c}{HIP~62207}\\
Parameter & 70~$\mu$m & 100~$\mu$m & 160~$\mu$m\\
\hline
A$_{\rm image}$ [\arcsec] & 10.8~$\pm$~0.2 & 12.1~$\pm$~0.2 & 15.0~$\pm$~0.5 \\
B$_{\rm image}$ [\arcsec] & \phantom{0}8.0~$\pm$~0.2 & \phantom{0}7.9~$\pm$~0.1 & 11.4~$\pm$~0.3 \\
PSF [\arcsec] & 5.6 & 6.8 & 11.3 \\
A$_{\rm disc}$ [\arcsec] & \phantom{0}7.4~$\pm$~0.3 & \phantom{0}7.0~$\pm$~0.2 & \phantom{0}6.6~$\pm$~0.6 \\
A$_{\rm disc}$ [au] & 128.8~$\pm$~5.2 & 121.8~$\pm$~3.5 & 114.8~$\pm$~10.4 \\
$\theta$ [\degr] & 109~$\pm$~3 & 111~$\pm$~2 & 116~$\pm$~2 \\
$i$ [\degr] &\multicolumn{3}{c}{56~$\pm$~10} \\
\hline
          & \multicolumn{3}{c}{HIP~72848}\\
Parameter & 70~$\mu$m & 100~$\mu$m & 160~$\mu$m\\
\hline
A$_{\rm image}$ [\arcsec] & \phantom{0}7.4~$\pm$~0.3 & 11.8~$\pm$~0.4 & 12.1~$\pm$~0.8 \\
B$_{\rm image}$ [\arcsec] & \phantom{0}5.6~$\pm$~0.2 & \phantom{0}6.9~$\pm$~0.2 & 10.7~$\pm$~1.3 \\
PSF [\arcsec] & 5.6 & 6.8 & 11.3 \\
A$_{\rm disc}$ [\arcsec] & \phantom{0}3.2~$\pm$~0.4 & \phantom{0}5.9~$\pm$~0.4 & \phantom{0}1.3~$\pm$~1.5 \\
A$_{\rm disc}$ [au] & \phantom{0}36.8~$\pm$~4.6 & \phantom{0}67.9~$\pm$~4.6 & \phantom{0}15.0~$\pm$~17.3 \\
$\theta$ [\degr] & \phantom{0}45~$\pm$~5 & \phantom{0}66~$\pm$~2 & \phantom{0}76~$\pm$~29 \\
$i$ [\degr] &\multicolumn{3}{c}{84~$\pm$~10}\\
\hline
\end{tabular}
\tablefoot{A$_{\rm image}$ = FWHM along source major axis; B$_{\rm image}$ = FWHM along source minor axis; PSF = average FWHM of instrument beam;\\ A$_{\rm disc}$ = deconvolved disc extent; $\theta$ = position angle; $i$ = inclination.}
\end{table}

\section{Modelling}  \label{sect_mod}

To ascertain the robustness of our results we have used three distinct approaches of increasing complexity to interpret the extended emission from these discs, namely a single annulus model, a modified black body model, and a radiative transfer model. We summarize each of these methods below.

\subsection{Single annulus}

In the first approach, we use the method of \cite{booth13} to interpret the extended emission. We model the disc structure as a simple annulus extending from its inner edge, $R_{\rm in}$, to a distance $R_{\rm in} + 0.1\times R_{\rm in}$. This approach is justified as we only see excess at far-infrared wavelengths which can be interpreted as cold dust at a single radial location; there is no evidence for warm excess emission at mid-infrared wavelengths for any of the three targets, which would imply the presence of a second physically distinct component to the debris disc architecture. The radial surface density profile within the annulus is assumed to be flat as the disc extent is narrow and this avoids assumptions regarding the radial distribution of material. Note that the fitting process adopted here ignores the disc SED or any realistic grain properties, i.e. it is a purely structural fit to the observed emission. 

Fitting is undertaken using a grid of models. Each individual model is defined by its inner radius, $R_{\rm in}$, position angle, $\theta$, and inclination to the line of sight, $i$. A line of sight integrator is used to create the model image which is then scaled to the observed flux density (using the least-squares minimization code, MPFIT) before the star's photospheric contribution is added to the centre of the model image. The resulting model is convolved with the instrument PSF of each PACS waveband to produce a set of three model images. To determine the best fit model parameters, their uncertainties and the source flux density, the reduced $\chi^{2}$ is calculated by comparison of the observed image to each grid model at each wavelength. All pixels within a distance of 1.5$\times$FWHM of the resolved disc (calculated from a 2D Gaussian fit) with a flux density greater than the 1-$\sigma$ r.m.s. are used in the fitting. The best fitting model is that which has the lowest reduced $\chi^{2}$. Uncertainties are determined by finding the maximum and minimum values of each parameter amongst models that lie within a $\Delta\chi^{2}$ equivalent to 1-$\sigma$ of the best fit model (3.53, in the case of a three parameter fit). In the case of contamination of the disc by a nearby background source (e.g. HIP~62207) we prevent the background source from influencing the fitting results by masking the region around the source from the $\chi^{2}$ calculation. 

\subsection{Modified blackbody}

In this approach we use the method outlined in \cite{kennedy12a} and \cite{wyatt12} to model the discs' emission combining the information from both the radial profiles and SED, but without adopting a 'realistic' grain size distribution or composition.

We model the disc as an annulus extending between radii $R_{\rm in}$ and $R_{\rm out}$. The disc radial profile was defined as a function of radius such that $\tau(R)\!=\!\tau_{0}(R/R_{0})^{\alpha}$; the emission from the annuli within the disc is determined by the vertical geometrical optical depth, $\tau(R)$, the geometric optical depth at the inner radius is $\tau_{0}$ and the slope of the radial density profile is $\alpha$. To avoid complications with the unconstrained dust composition and emissivity, the thermal emission of the disc at a given radius is assumed to be well approximated by a modified blackbody defined by a temperature, $T$, and break wavelength, $\lambda_{0}$, beyond which the spectral slope of the emission is modified by a factor $\beta$ such that $F\!=\!F\times(\lambda_{0}/\lambda)^{\beta}$ for $\lambda > \lambda_{0}$. The temperature is assumed to be $T\!=\!f_{T}T_{\rm bb}$, where $T_{\rm bb}$ is the blackbody temperature at radius $R$ and $f_{T}$ is a free parameter. An image of the disc is produced by integrating the emission along the line of sight to the observer given the orientation of the disc, given by the position angle, $\theta$, and inclination, $i$. The appropriate stellar photosphere contribution is added before convolution with the PSF to create a set of model images to be compared to the observations. 

To quantify the success of a model in replicating the observed emission, a $\chi^{2}$ minimization approach was undertaken. A $\chi^{2}$ is derived from the SED as well as the images (which were calculated using the method above), but the SED $\chi^{2}$ only included photometry which was not taken from the \textit{Herschel} images. These $\chi^{2}$ values were then summed. For the results presented here, the fitting process was undertaken both by hand and by least-squares minimization, so the search of parameter space was neither exhastive, nor systematic, and therefore uncertainties are not given. However, we note the 'best fit' models presented in Table \ref{mod_table} appear to lie in the global minimum of the total parameter space and are in good agreement with the other, systematic approaches. 

\subsection{{\sc GRaTeR}}

Finally, we use {\sc GRaTeR}, a radiative transfer debris disc modelling code \citep{augereau99,lebreton12}, to interpret the observations. As input to the modelling process, we take the available \textit{WISE}, \textit{Spitzer} MIPS and IRS photometry, along with the new \textit{Herschel} PACS and SPIRE photometry to produce an SED for each target. The spatial extent of the disc is represented by the radial profiles from the three PACS bands, with the instrument PSF again represented by the profile of the bright standard star $\alpha$ Bo\"otis. The stellar contribution to the total flux density is taken from the fitted PHOENIX/GAIA model photosphere. The disc is fitted by a power-law model across a grid of parameters. The dust particle size distribution slope, $\gamma$, samples 13 values between 5.1 and 2.7 in steps of 0.2. The minimum grain size, $s_{\rm min}$, is sampled in 55 steps logaritmically spaced between 0.05 and 43.5~$\mu$m. The largest grain size, $s_{\rm max}$, is fixed at 1~mm due to the decreasing contribution of larger grains to the observed flux density at far-infrared wavelengths. The dust grain composition is assumed to be pure astronomical silicate \citep{draine03}, with the optical constants computed through Mie theory.  We characterize the radial component of the disc surface density profile, $R(r)$,  using a peak radius, $r_{0}$ and a two component power law of the form:

$R(r) \propto \left(\left(\frac{r}{r_{0}}\right)^{-2\alpha_{\rm in}} + \left(\frac{r}{r_{0}}\right)^{-2\alpha_{\rm out}}\right)^{-\frac{1}{2}}$

\noindent for the radial distribution of the dust grains, where $r$ is radial distance from the star, $r_{0}$ is the peak of the surface density profile whilst $\alpha_{\rm in}$ and $\alpha_{\rm out}$ are the exponents of the surface density profile interior and exterior to $r_{0}$, respectively. This produces a smooth disc model with a characteristic peak radius for its surface density, representative of the morphology of isolated, single component debris discs. The peak of the disc surface density profile, $r_{0}$, is sampled at 35 values between 10 and 120~au, and the maximum disc extent fixed to 200~au. Interior to $r_{0}$, the surface density is forced to fall off very steeply, with $\alpha_{\rm in}$ fixed as 5. Exterior to $r_{0}$ the radial surface density slope, $\alpha_{\rm out}$ (hereafter referred to as $\alpha$), varies between 0.0 and -5.0 in steps of 0.5. The disc orientation and inclination were fixed to the values measured from the 2D Gaussian fit to the source in the PACS 100~$\mu$m image, since that band provides the best compromise between angular resolution, S/N and sensitivity to cold extendend emission in the available data set. Each image (i.e. pair of radial profiles) is given the same weighting as the target SED in the fitting process. 

To calculate the best fit model and probability distributions of the parameters for each disc we use a statistical (Bayesian) inference method. {\sc GRaTeR} produces a grid of models comprising the parameters detailed above with the number of degrees of freedom being 67, 40 and 35 for HIP~22263, HIP~62207 and HIP~72848, respectively. The simultaneous best fit model is the point in the calculated grid with the lowest $\chi^{2}$. The best fit values and uncertainties for individual parameters are derived from probability distribution functions, providing an advantage of {\sc GRaTeR} over the modified blackbody modelling process. From the $\chi^{2}$ values we can obtain the probability of the data given the model parameters, assuming the model is appropriate for the data and the uncertainties of the measurements are normally distributed. No particular value of any parameter is favoured in this analysis, so the weighting is uniform for each set of parameters. If the model is a good representation of the data, we would expect that the simultaneous best fit model and the values for individual parameters will be in close agreement. This would also mean that there is little or no correlation between parameters and/or their uncertainties. For further details of the fitting process see \cite{lebreton12}.

We also fit the data with SAnD \citep{ertel12,ertel14} in order to validate the {\sc GRaTeR} results. The results from both codes agree well which is why we only show the {\sc GRaTeR} results here.

\begin{figure*}[th!!]
\centering
\subfigure{\includegraphics[width=0.33\textwidth]{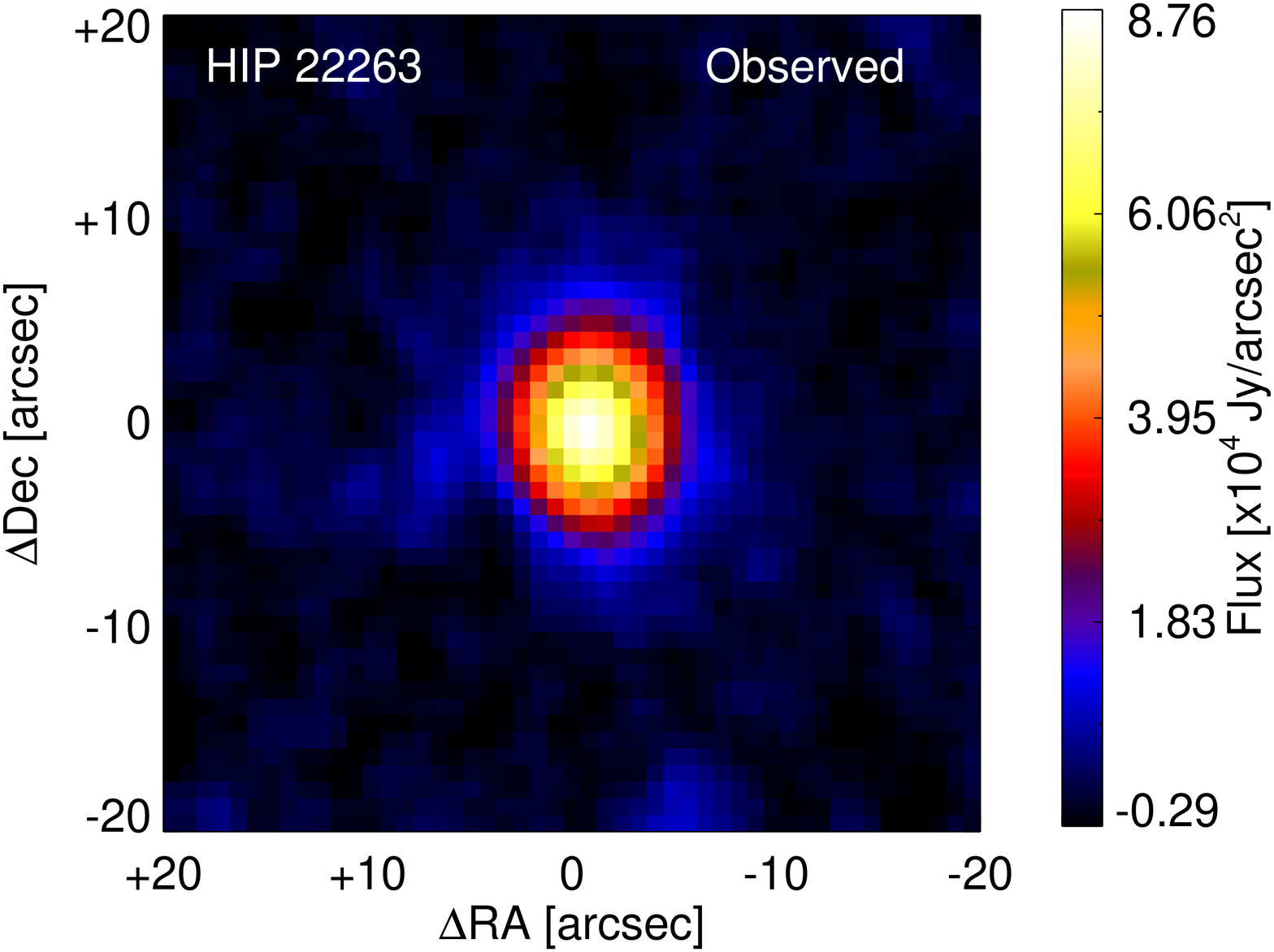}}
\subfigure{\includegraphics[width=0.33\textwidth]{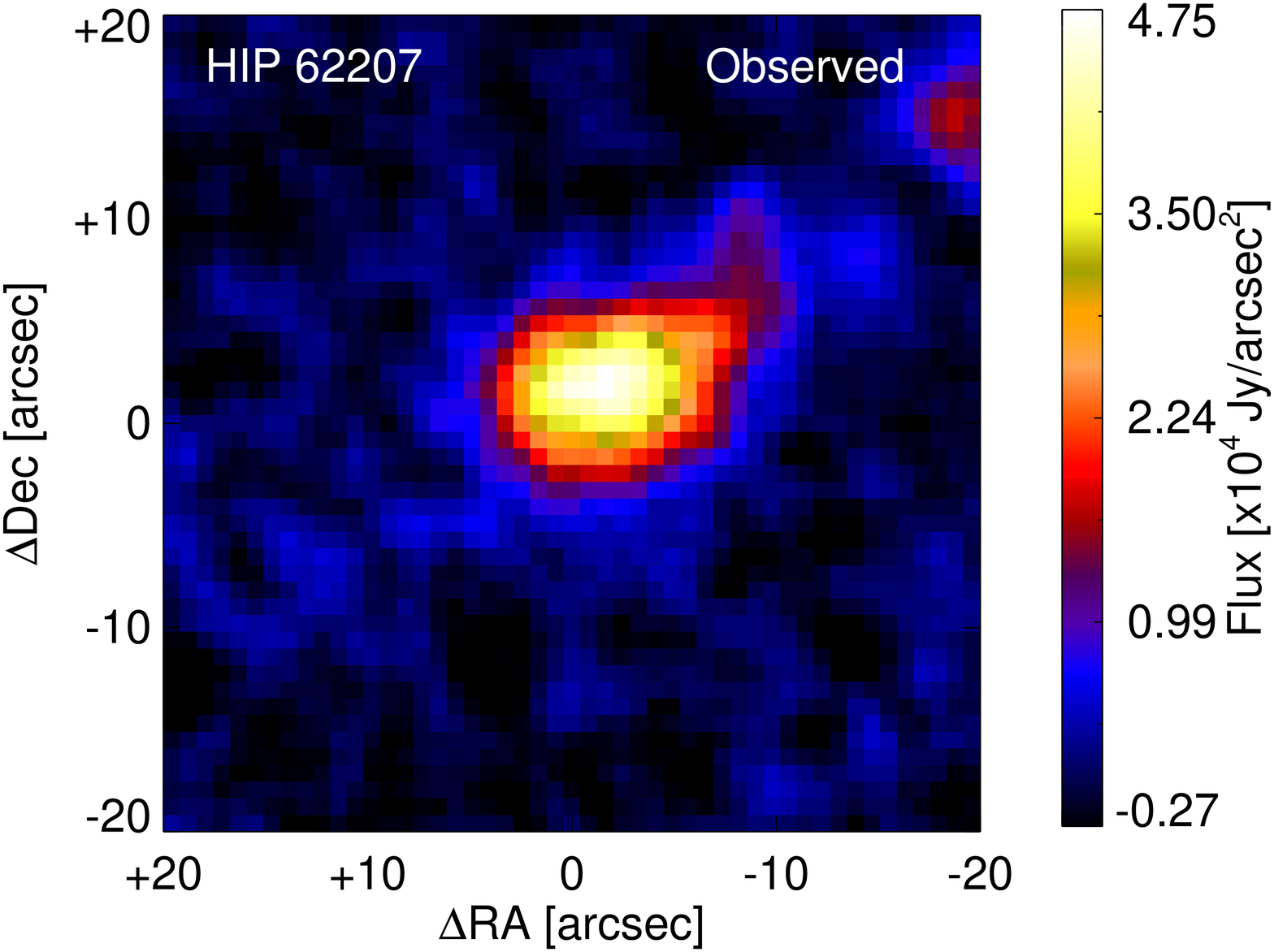}}
\subfigure{\includegraphics[width=0.33\textwidth]{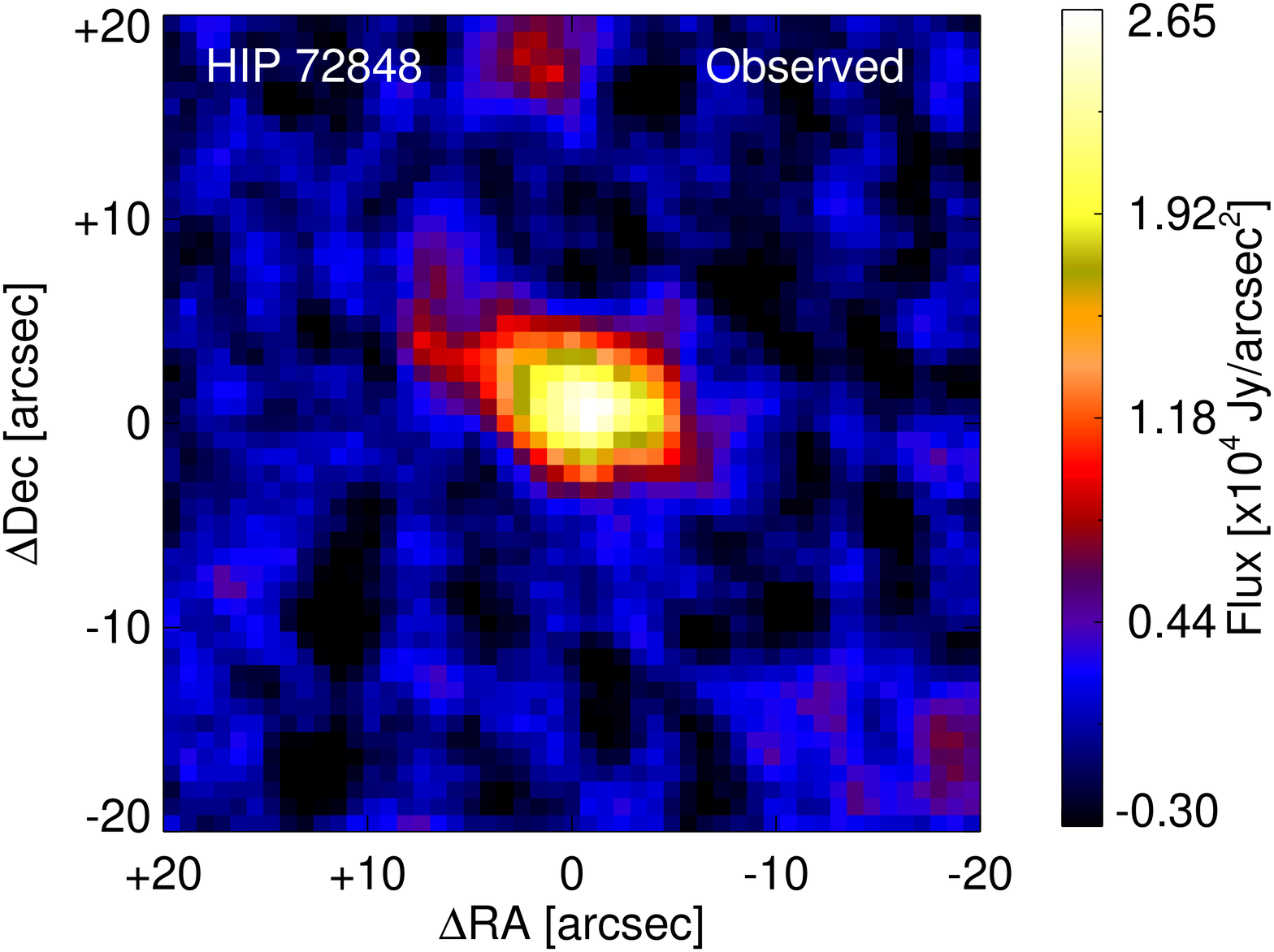}}
\subfigure{\includegraphics[width=0.33\textwidth]{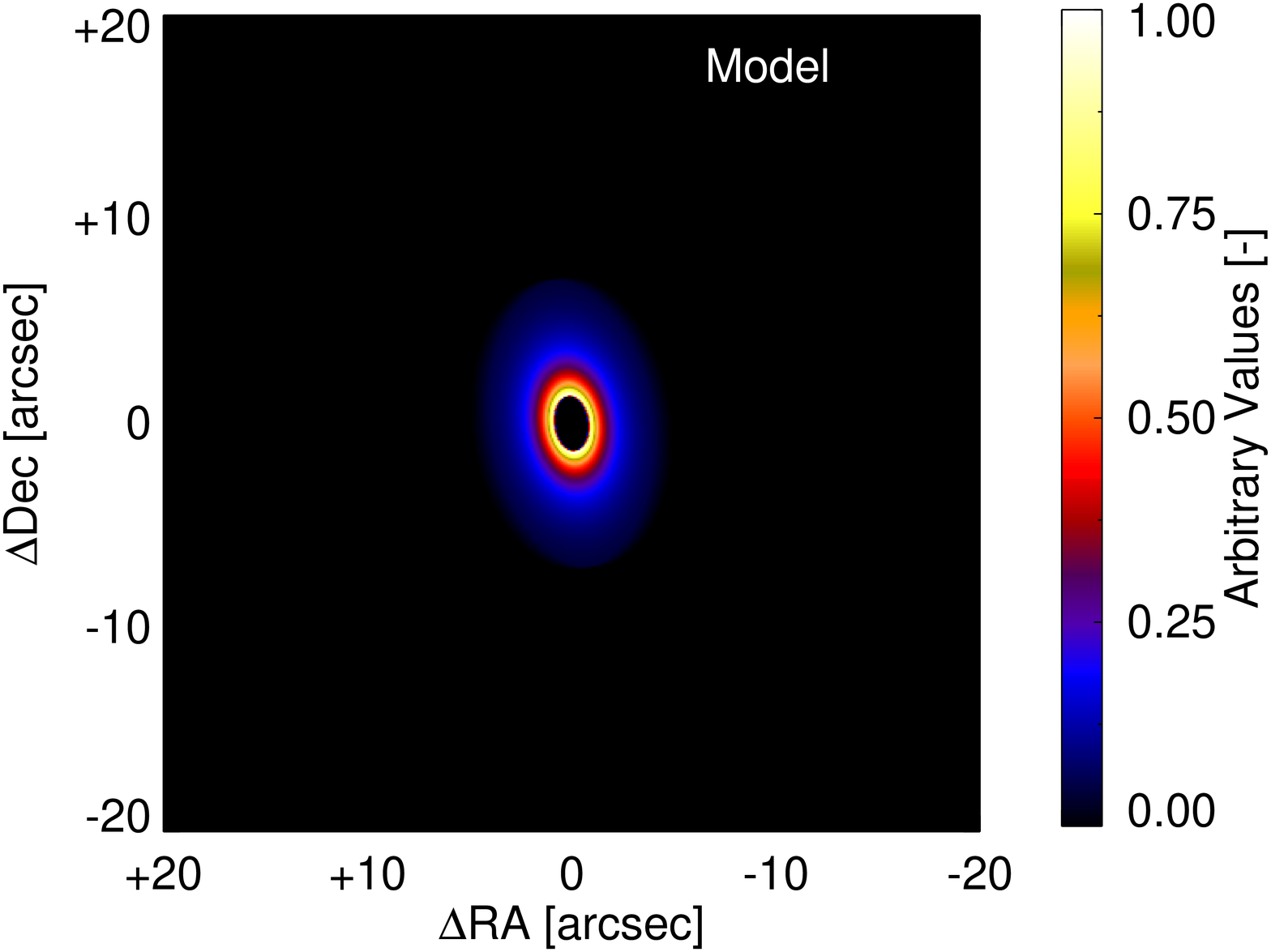}}
\subfigure{\includegraphics[width=0.33\textwidth]{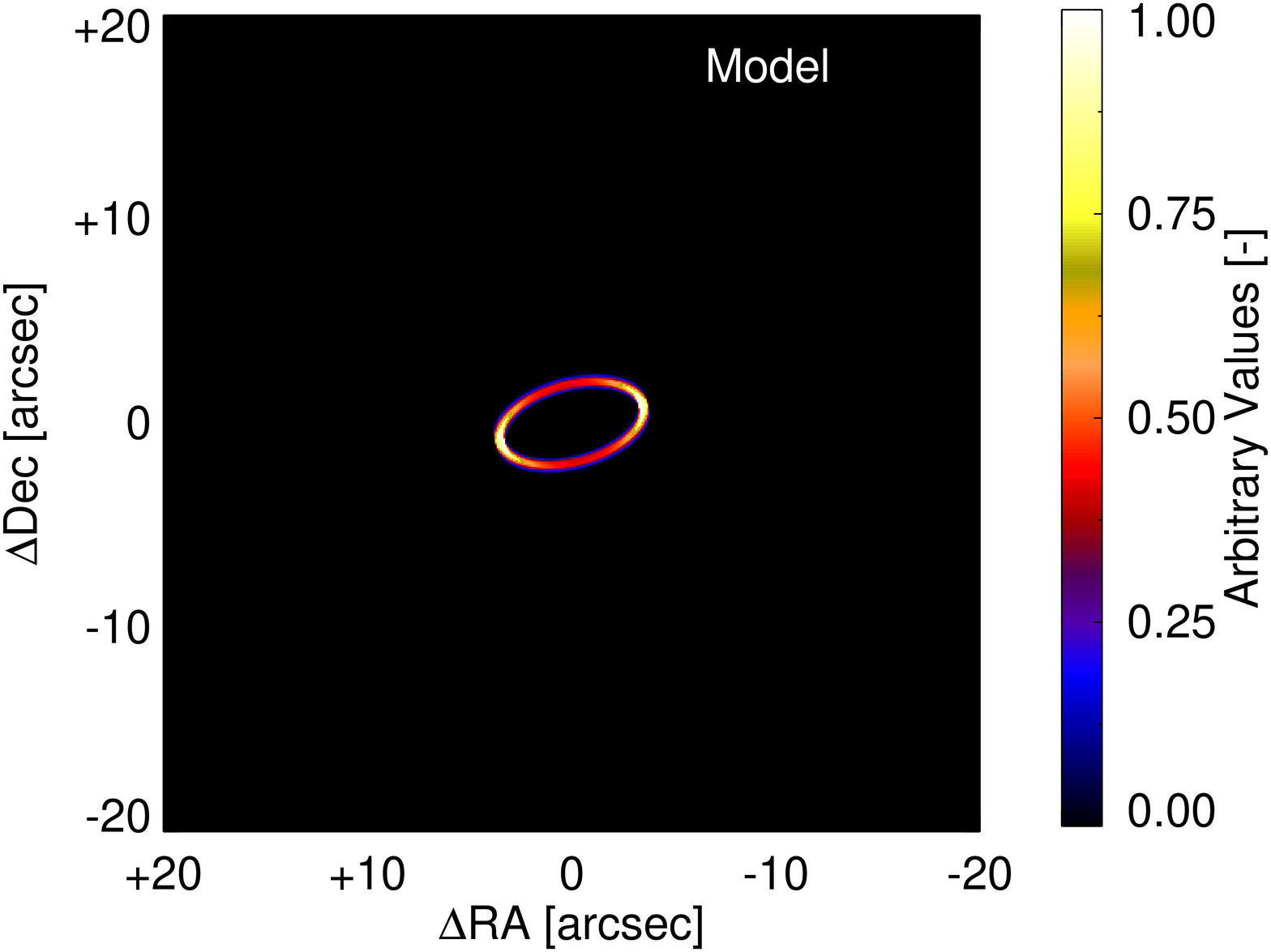}}
\subfigure{\includegraphics[width=0.33\textwidth]{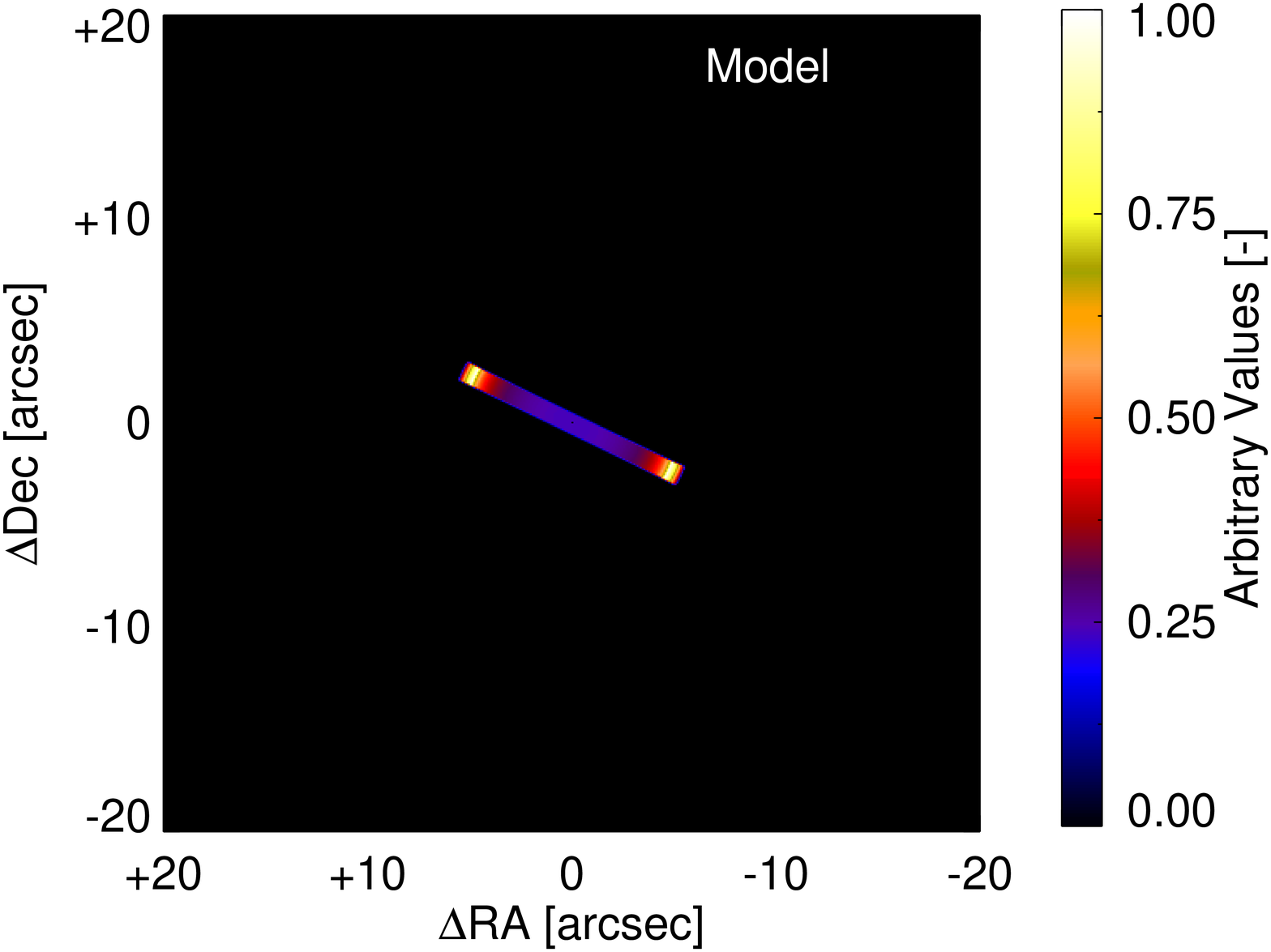}}
\subfigure{\includegraphics[width=0.33\textwidth]{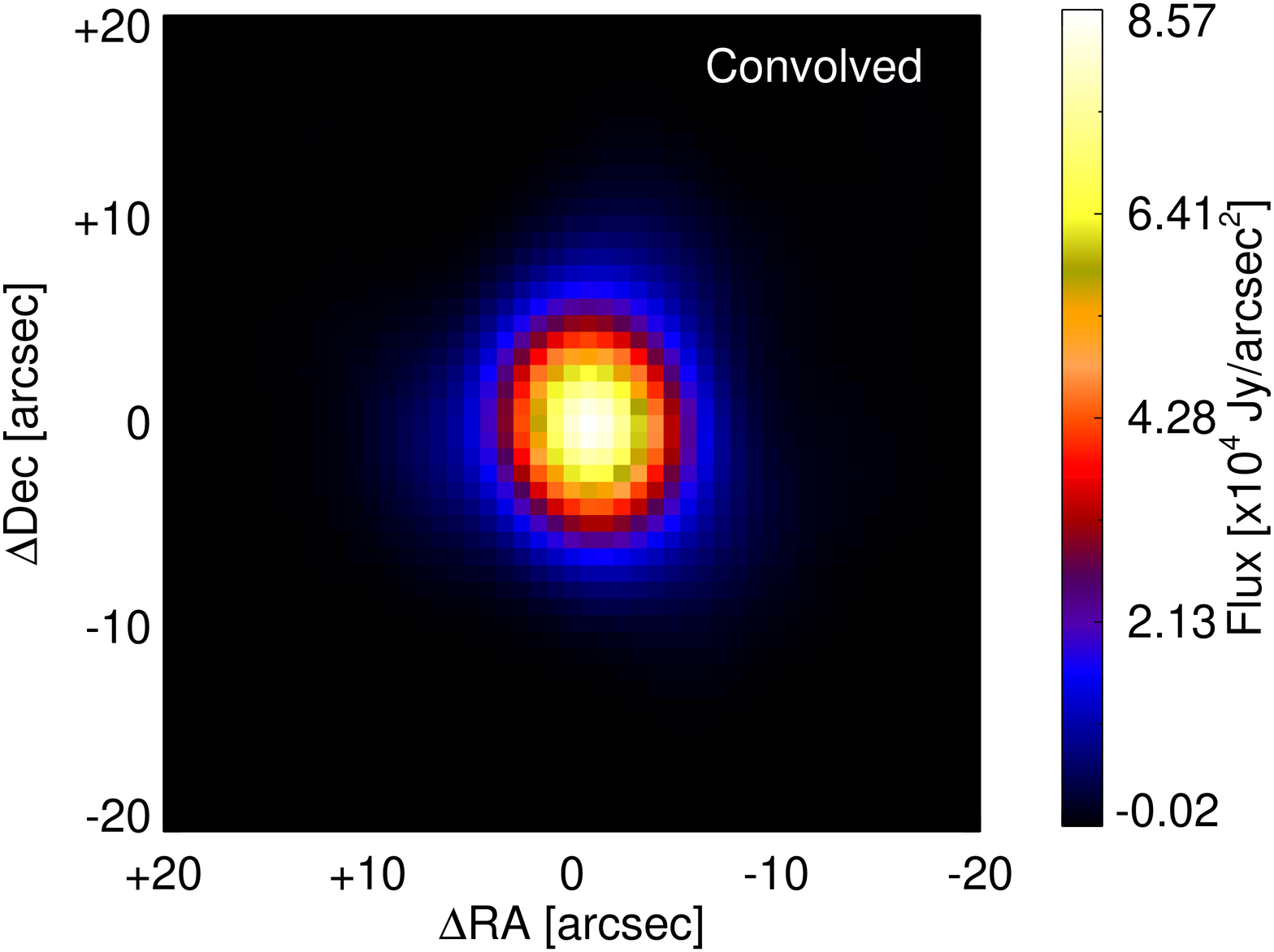}}
\subfigure{\includegraphics[width=0.33\textwidth]{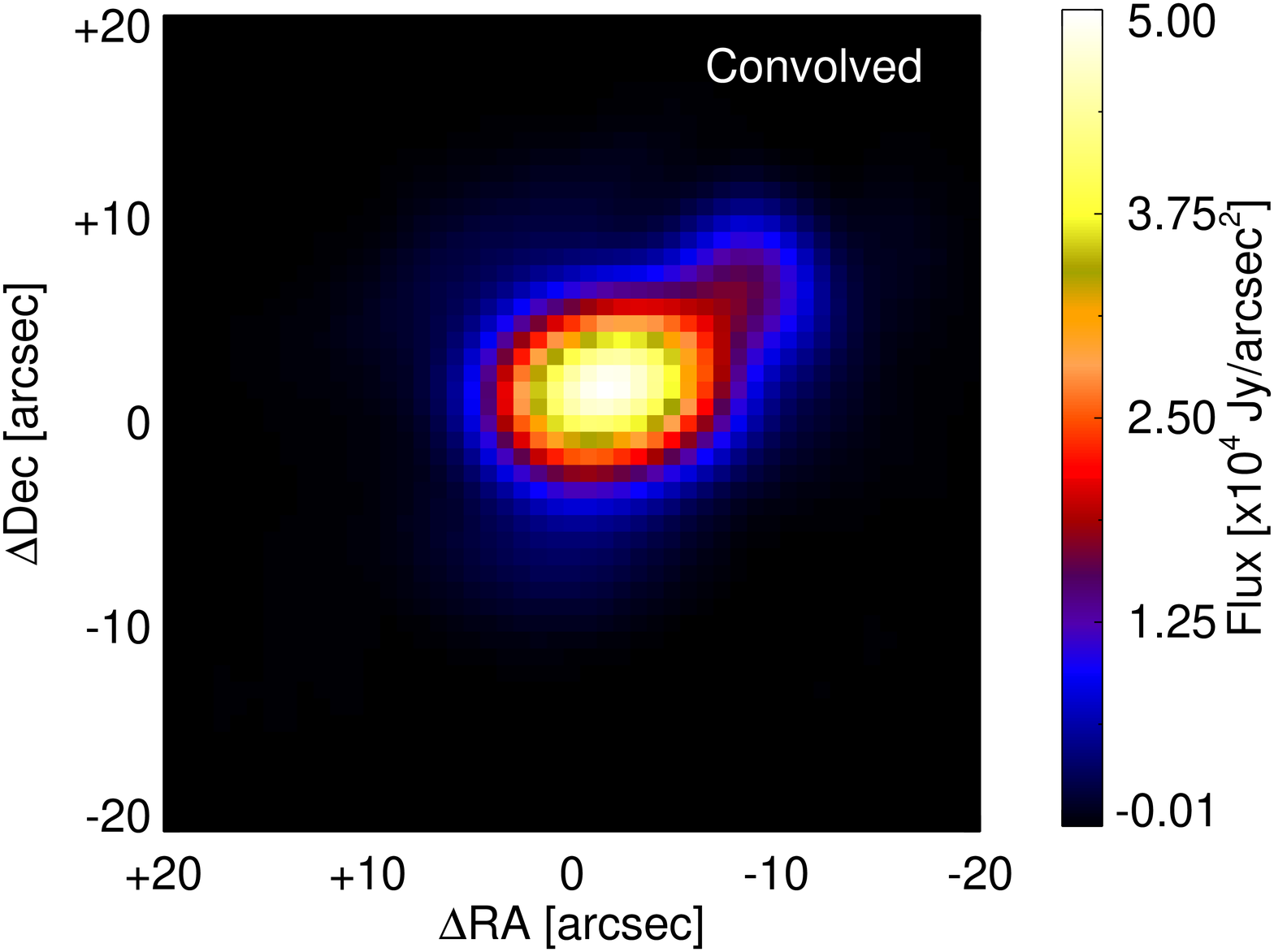}}
\subfigure{\includegraphics[width=0.33\textwidth]{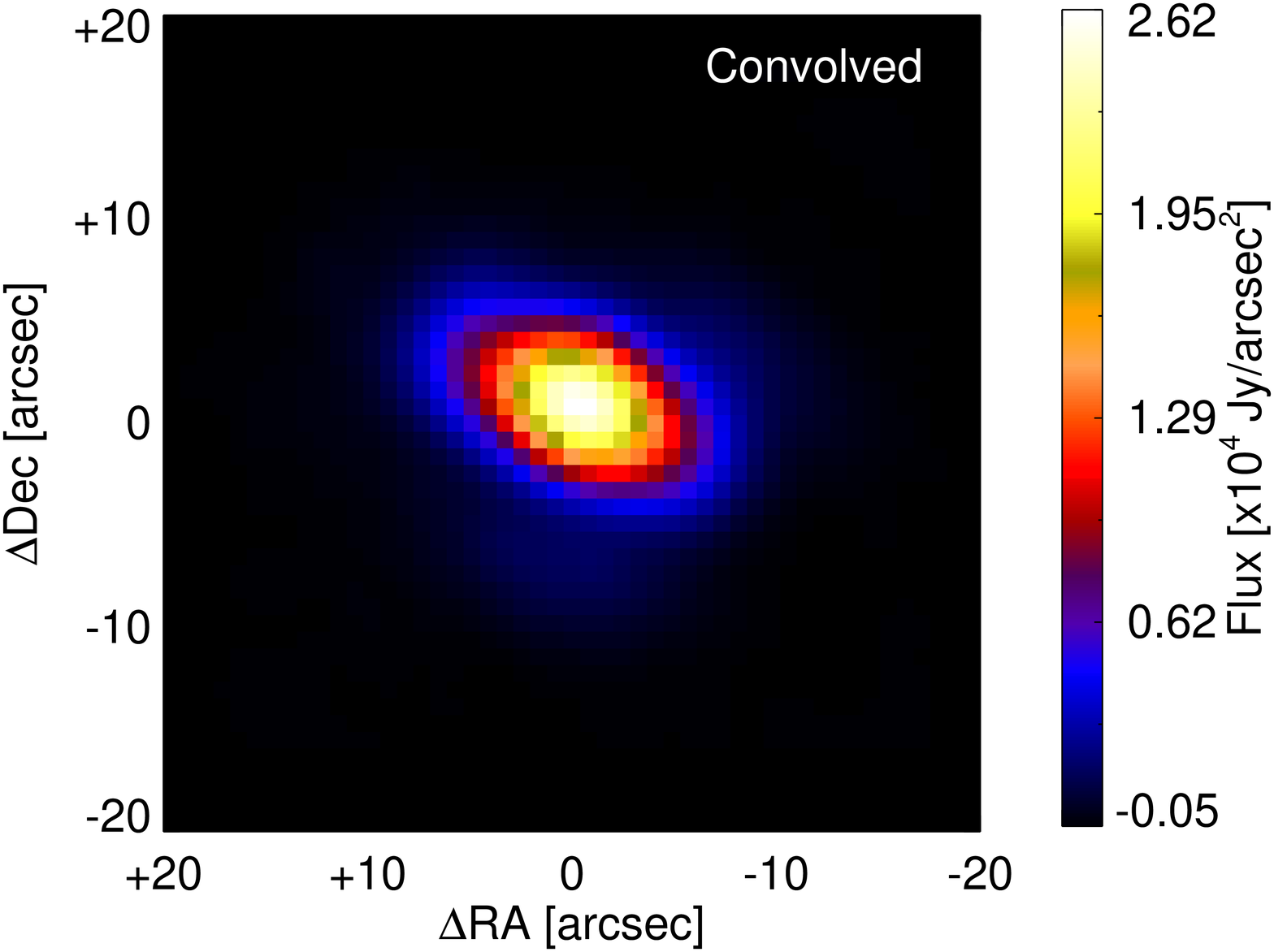}}
\subfigure{\includegraphics[width=0.33\textwidth]{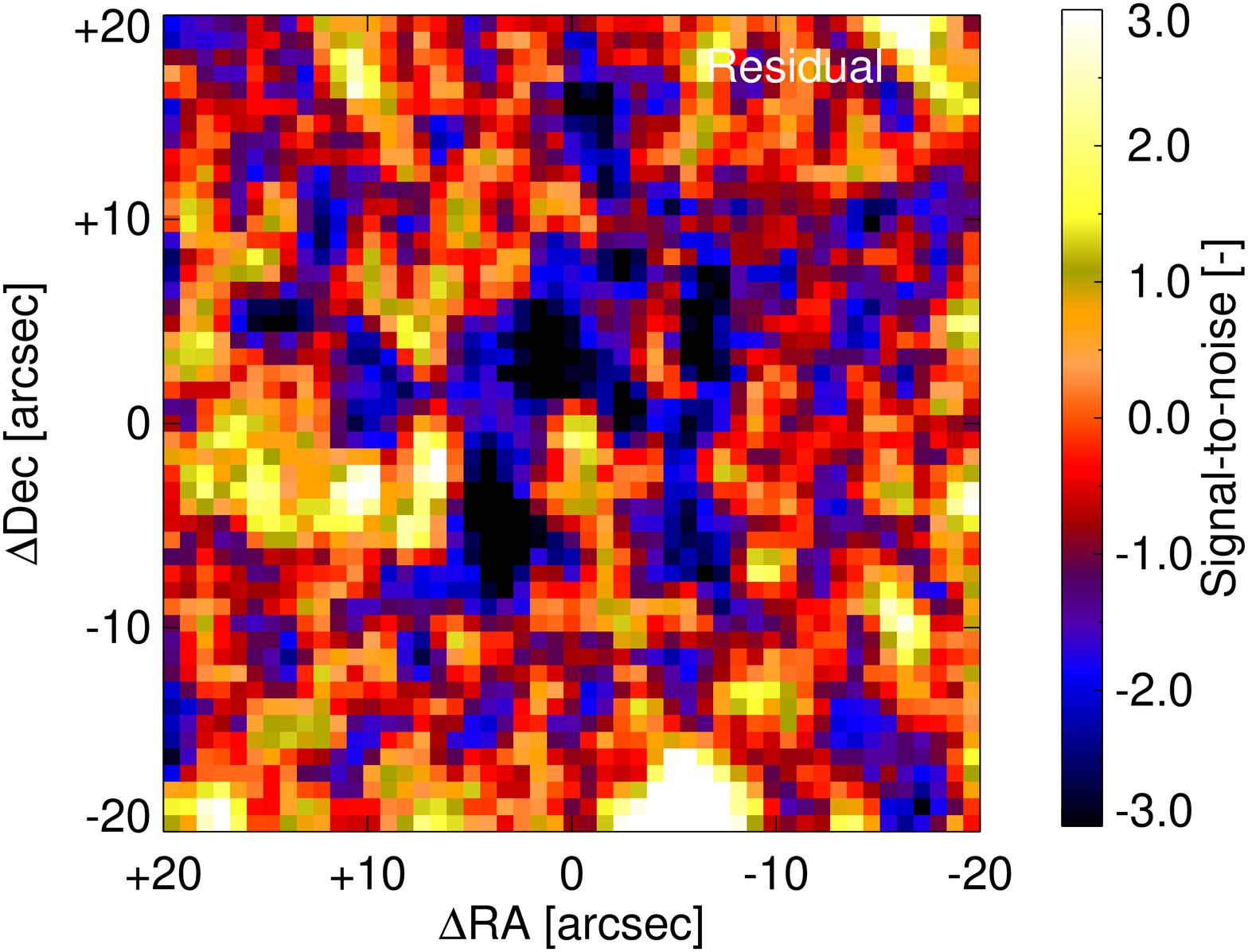}}
\subfigure{\includegraphics[width=0.33\textwidth]{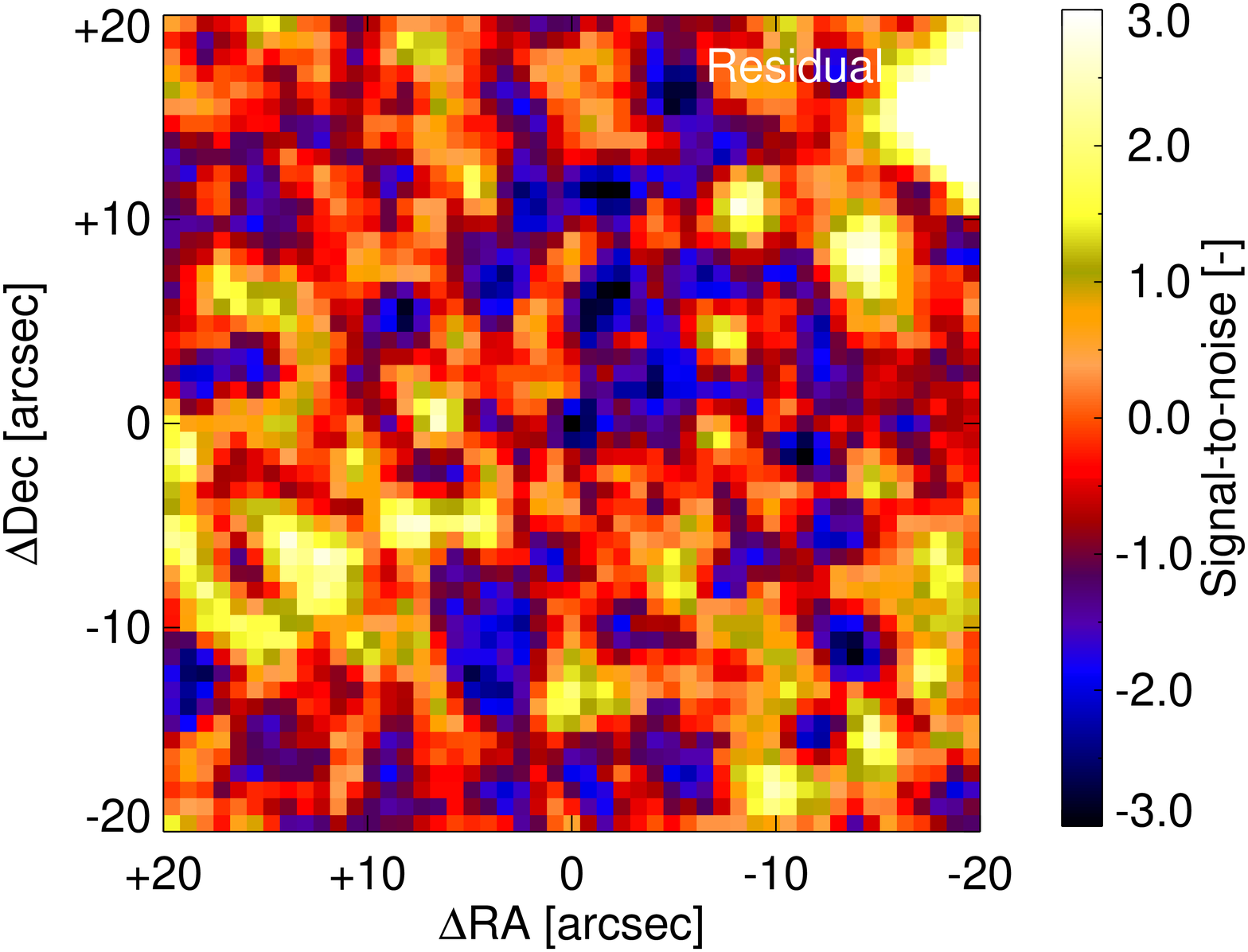}}
\subfigure{\includegraphics[width=0.33\textwidth]{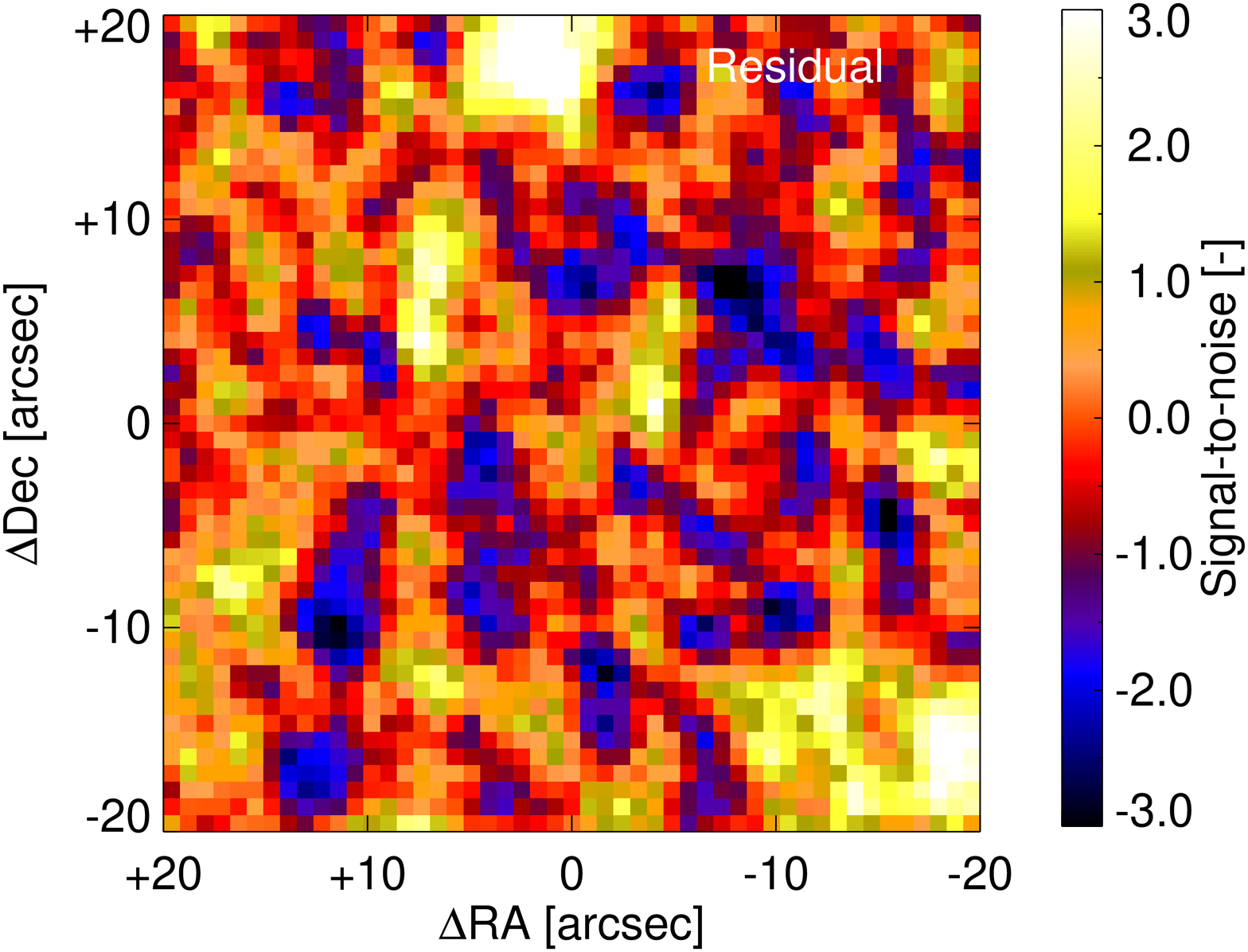}}
\caption{Modified blackbody results: Disc models of HIP~22263, HIP~62207 and HIP~72848 (left to right) at 100~$\mu$m. For each target the observation, ring model, convolved model (including star) and residual image (observed - convolved) are shown (top to bottom). Scalebar units are $\times10^{4}$ Jy/arcsec$^{2}$ for the observed and convolved images, but in units of signal-to-noise for the residual image. The HIP~62207 convolved image includes a background source not shown in the disc model image. Orientation is north up, east left and the image pixel scale is 1\arcsec.}
\label{mbb_plots}
\end{figure*}

\section{Results and discussion}  \label{sect_res}

\begin{figure*}[th!!]
\centering
\subfigure{\includegraphics[width=0.33\textwidth]{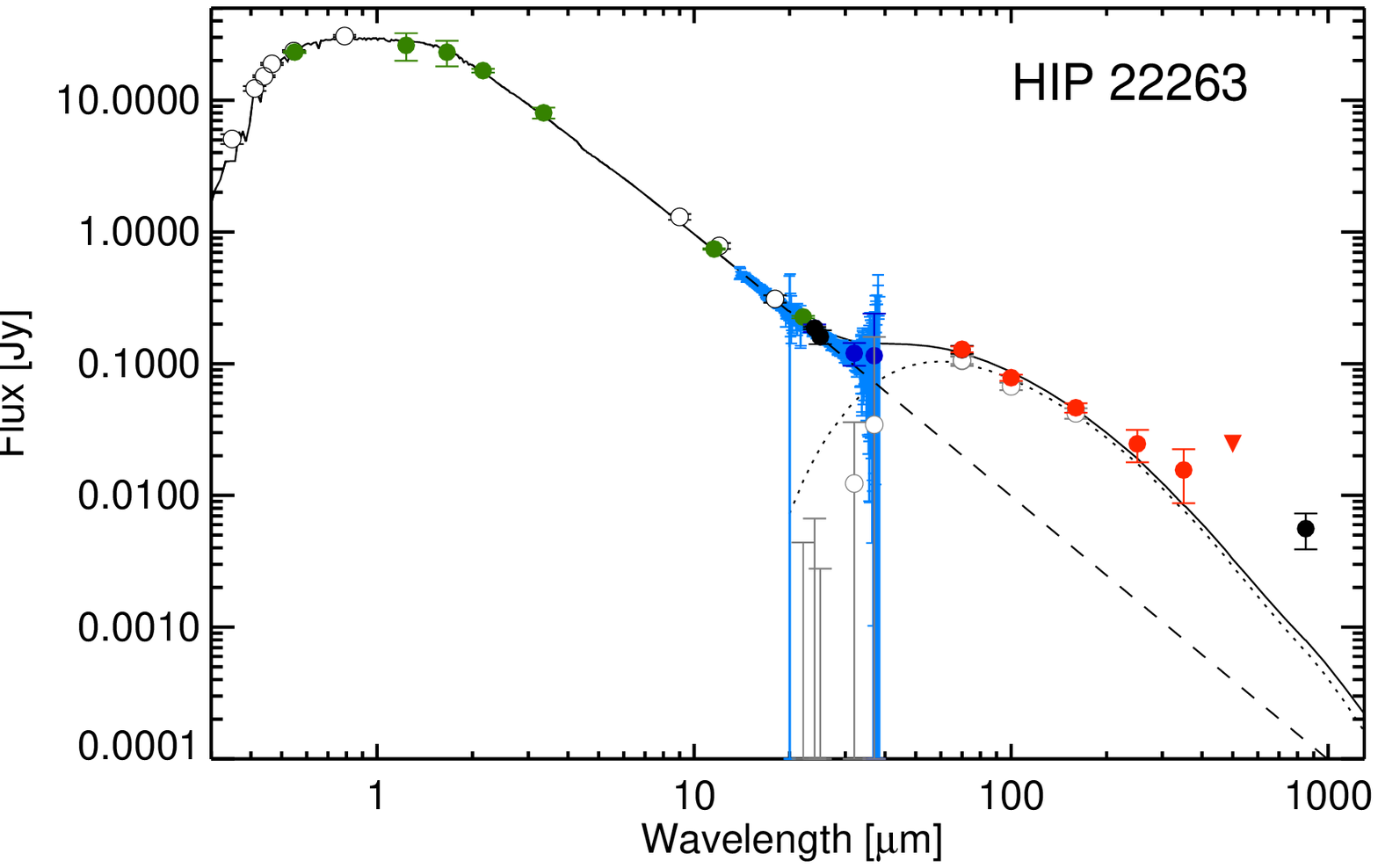}}
\subfigure{\includegraphics[width=0.33\textwidth]{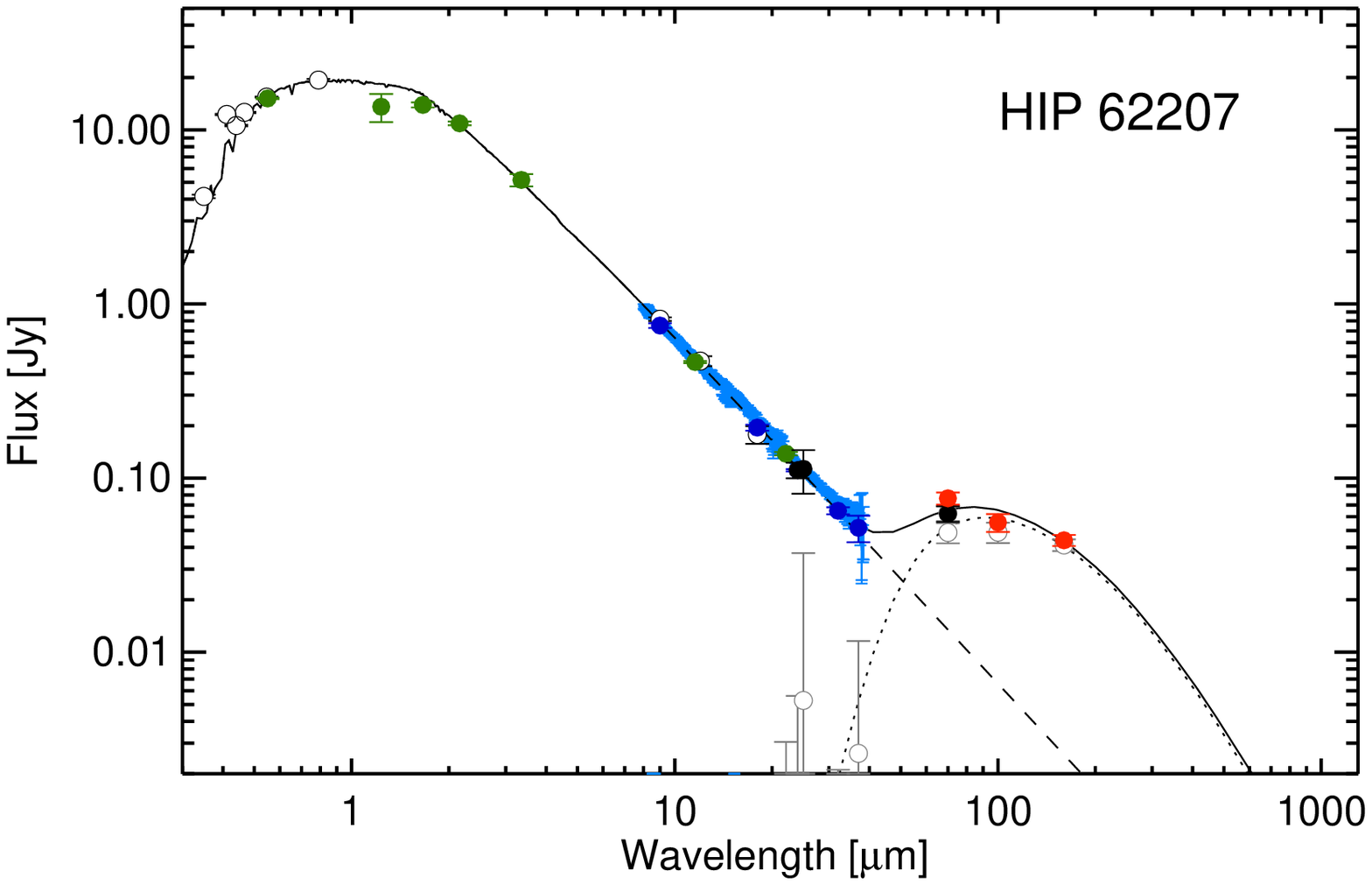}}
\subfigure{\includegraphics[width=0.33\textwidth]{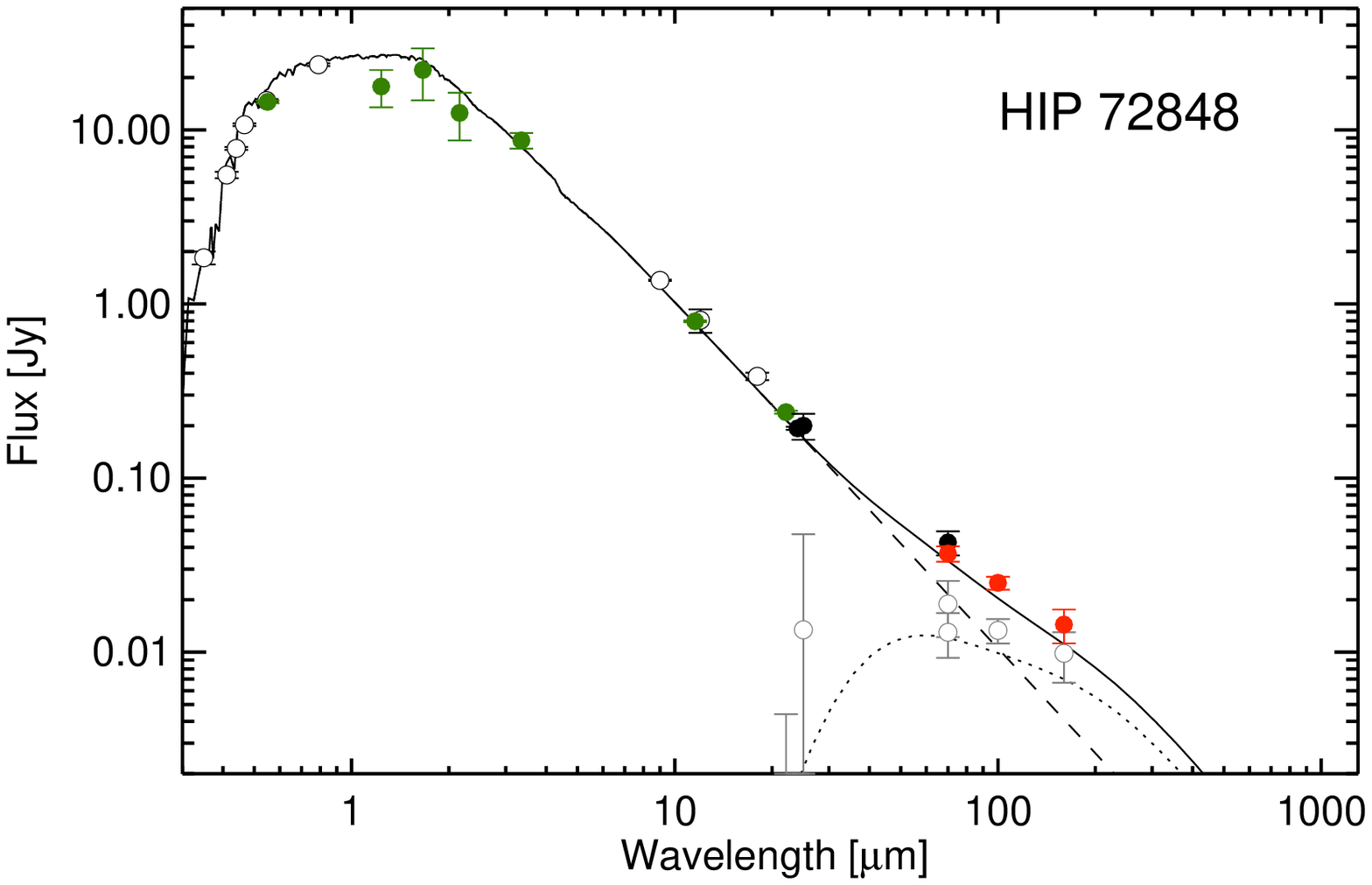}}
\subfigure{\includegraphics[width=0.33\textwidth]{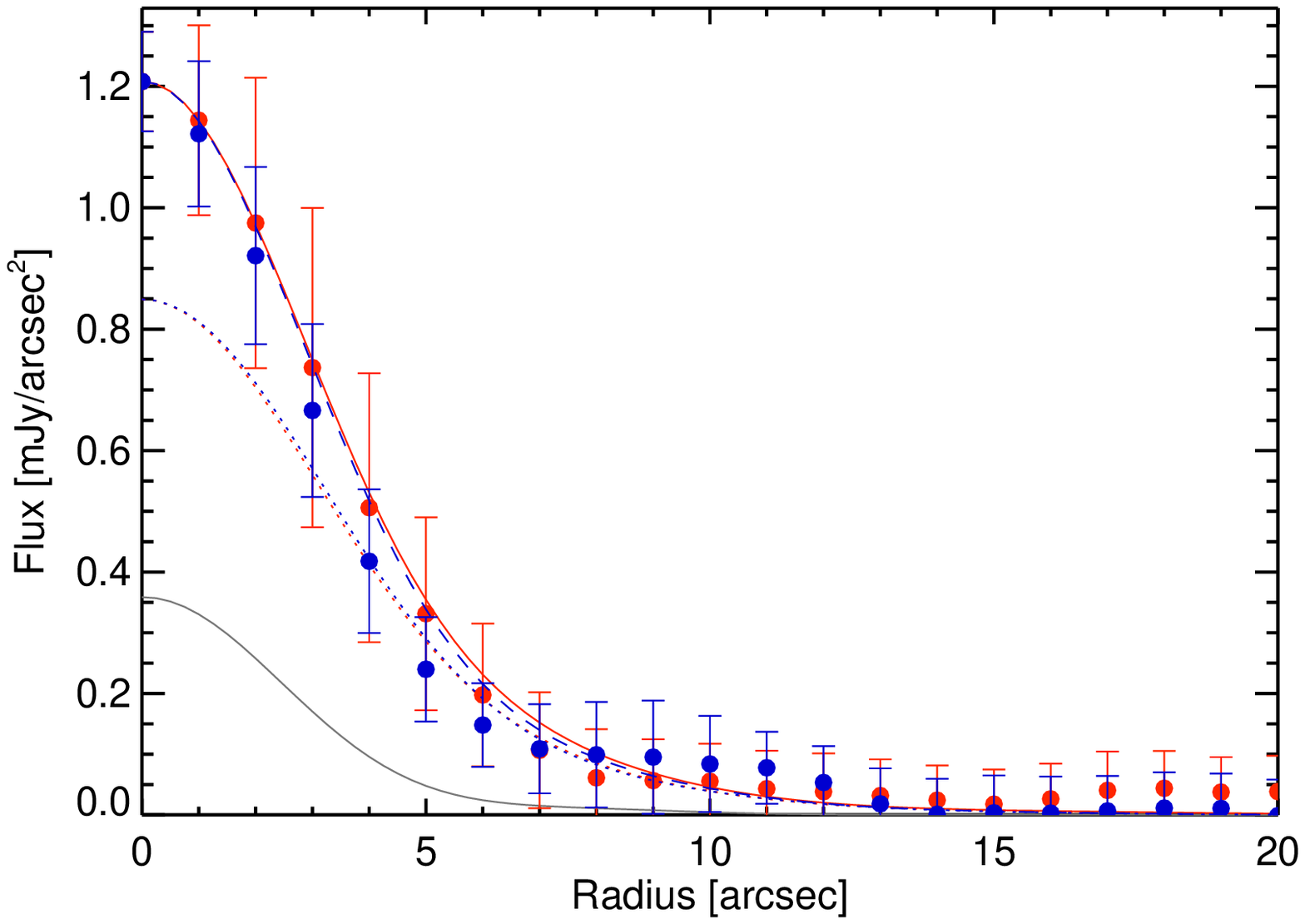}}
\subfigure{\includegraphics[width=0.33\textwidth]{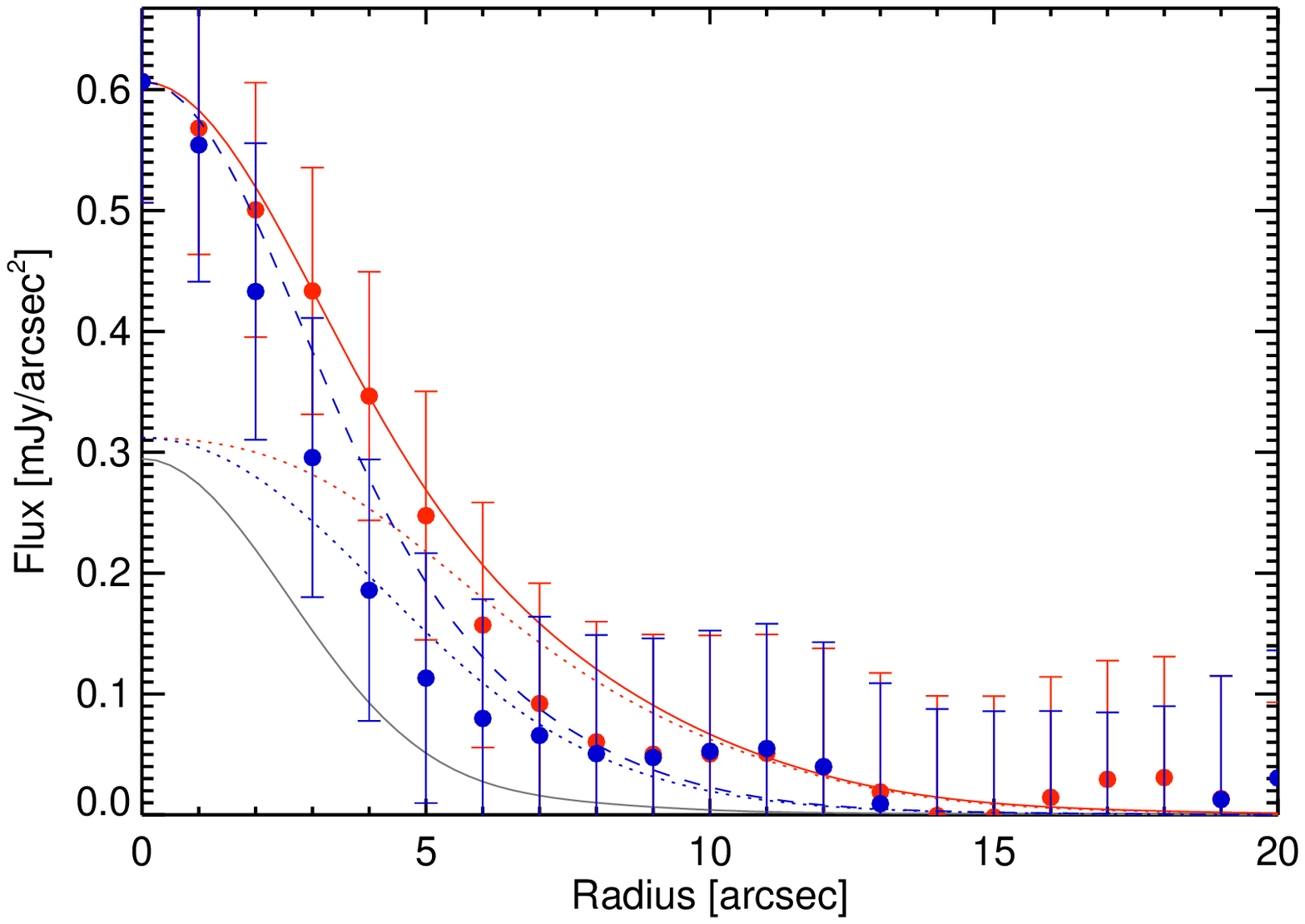}}
\subfigure{\includegraphics[width=0.33\textwidth]{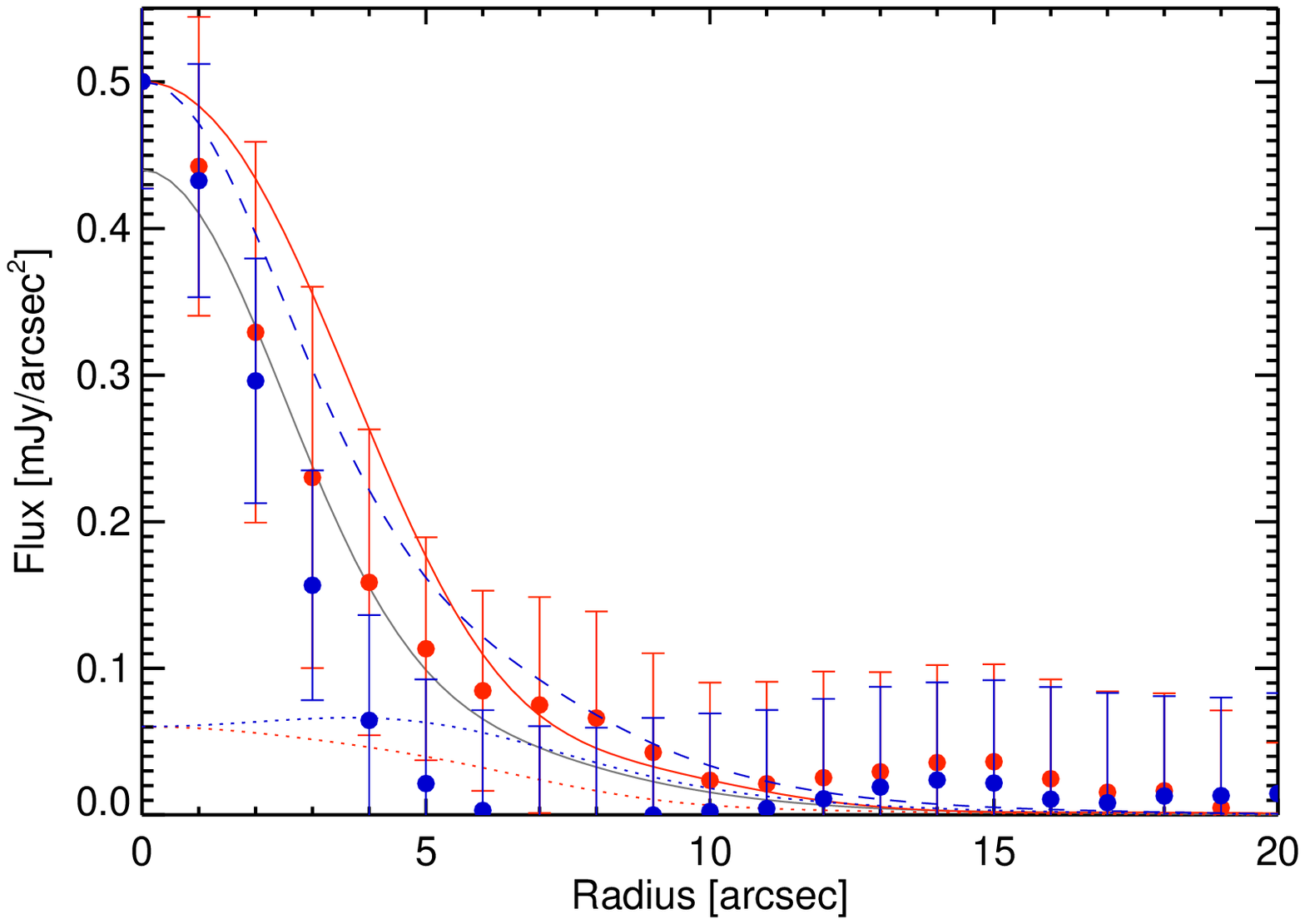}}
\subfigure{\includegraphics[width=0.33\textwidth]{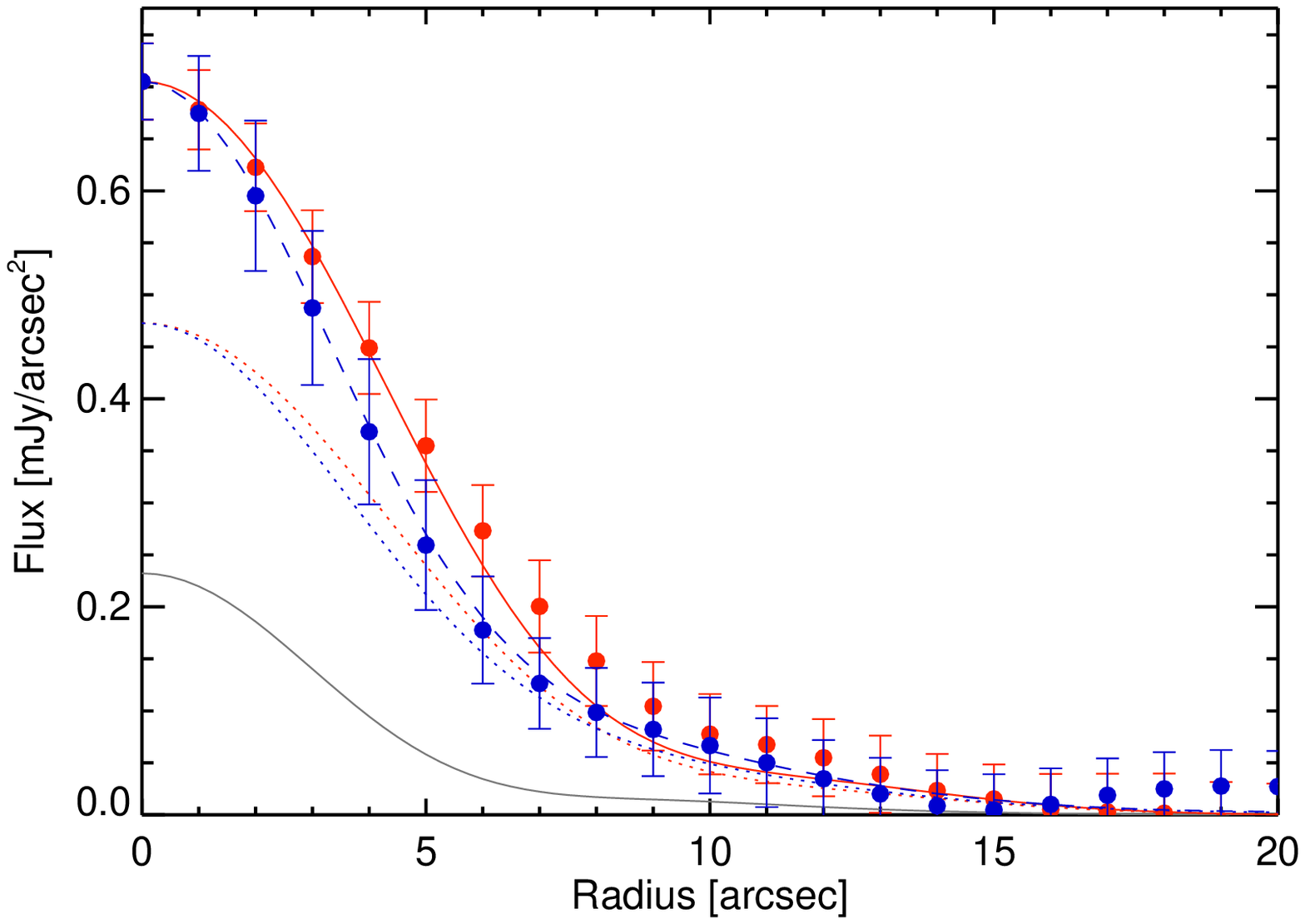}}
\subfigure{\includegraphics[width=0.33\textwidth]{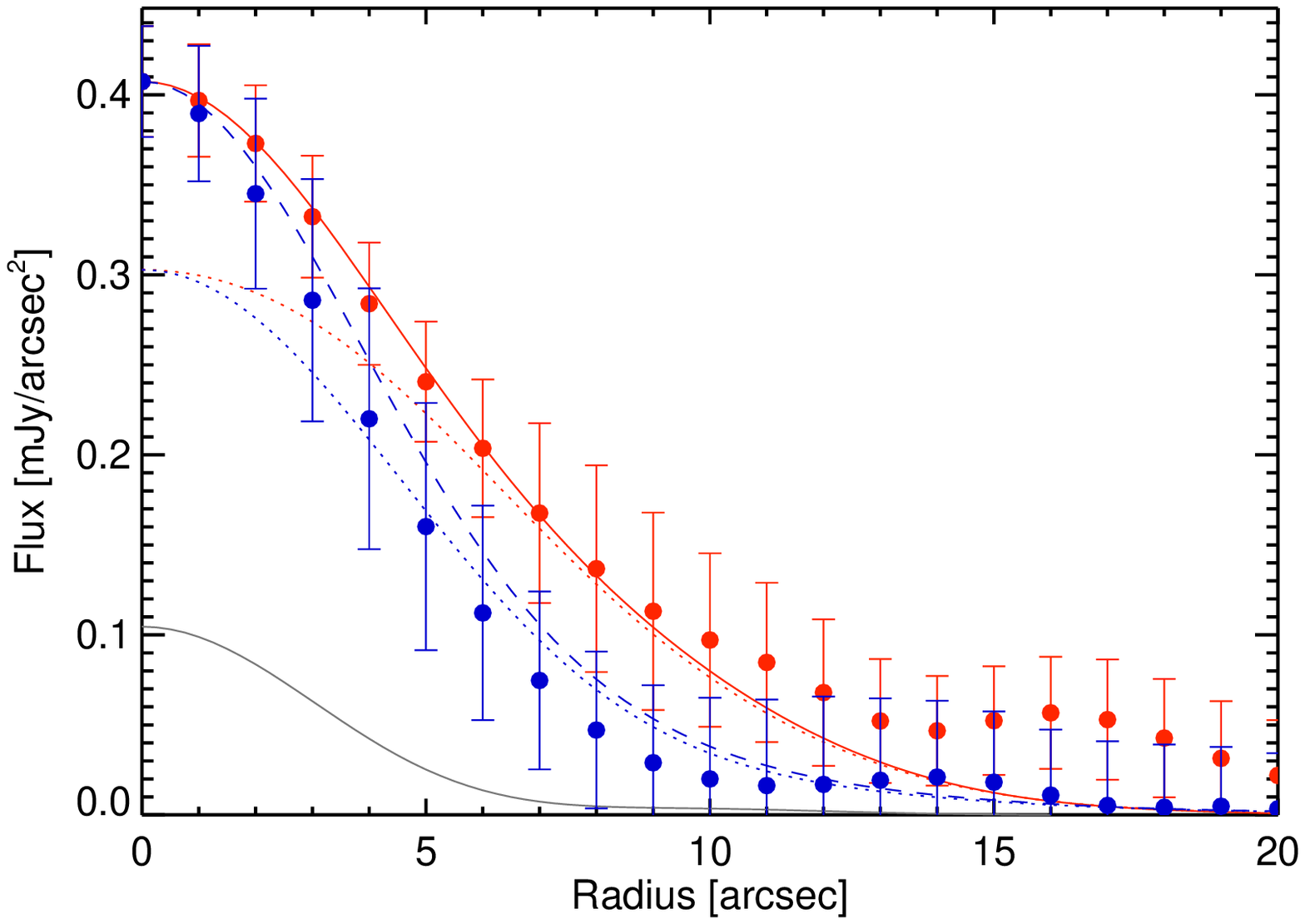}}
\subfigure{\includegraphics[width=0.33\textwidth]{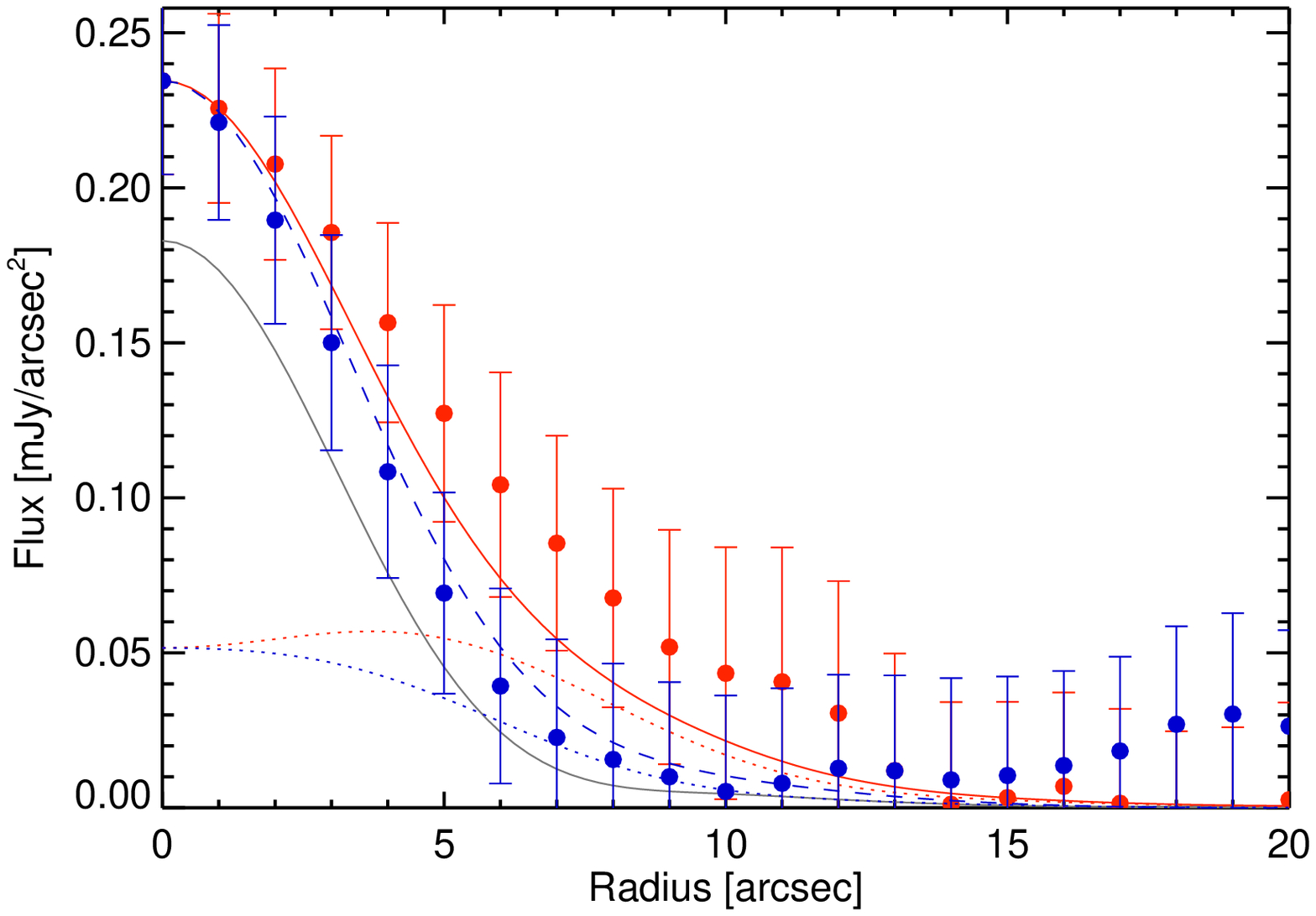}}
\subfigure{\includegraphics[width=0.33\textwidth]{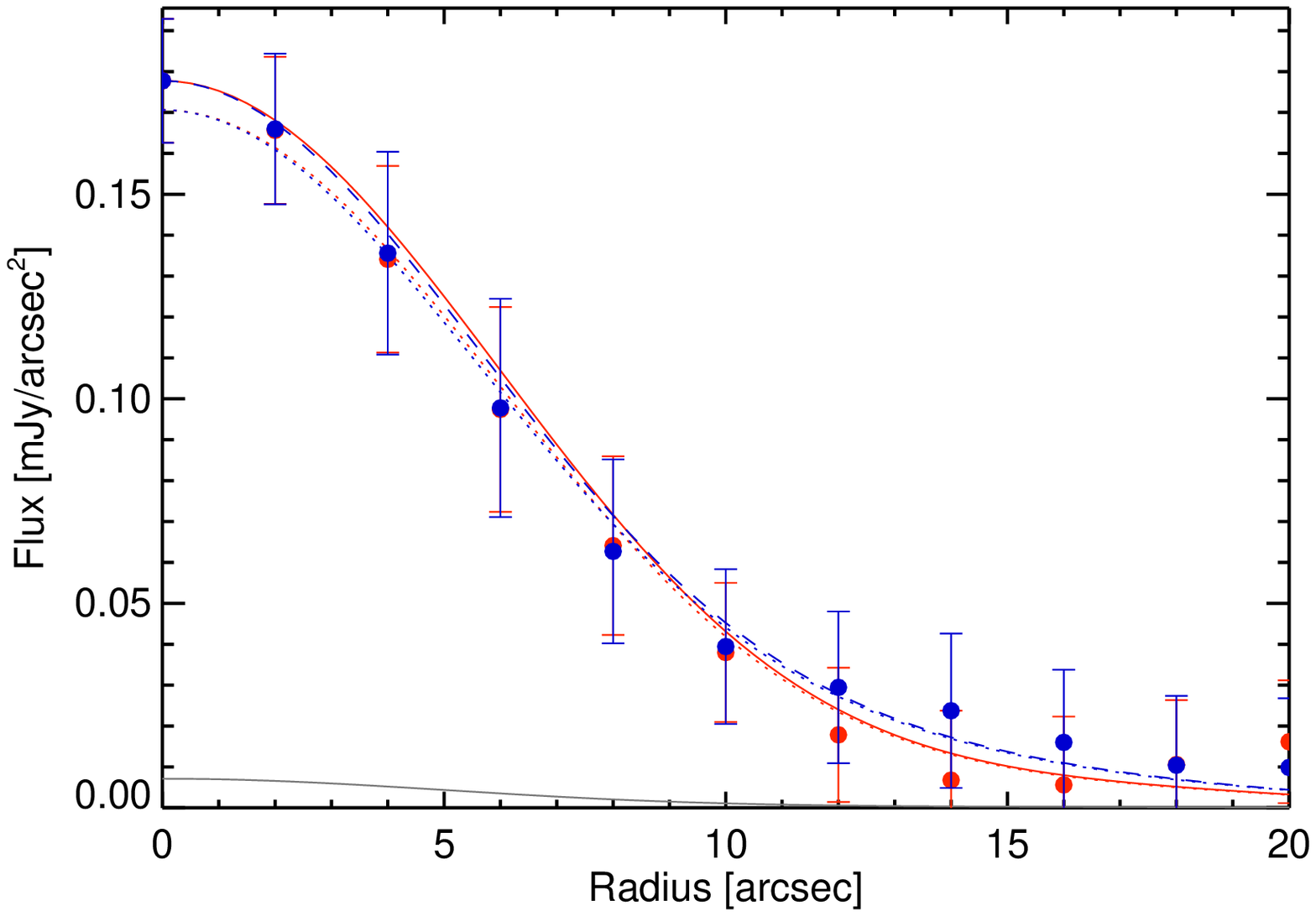}}
\subfigure{\includegraphics[width=0.33\textwidth]{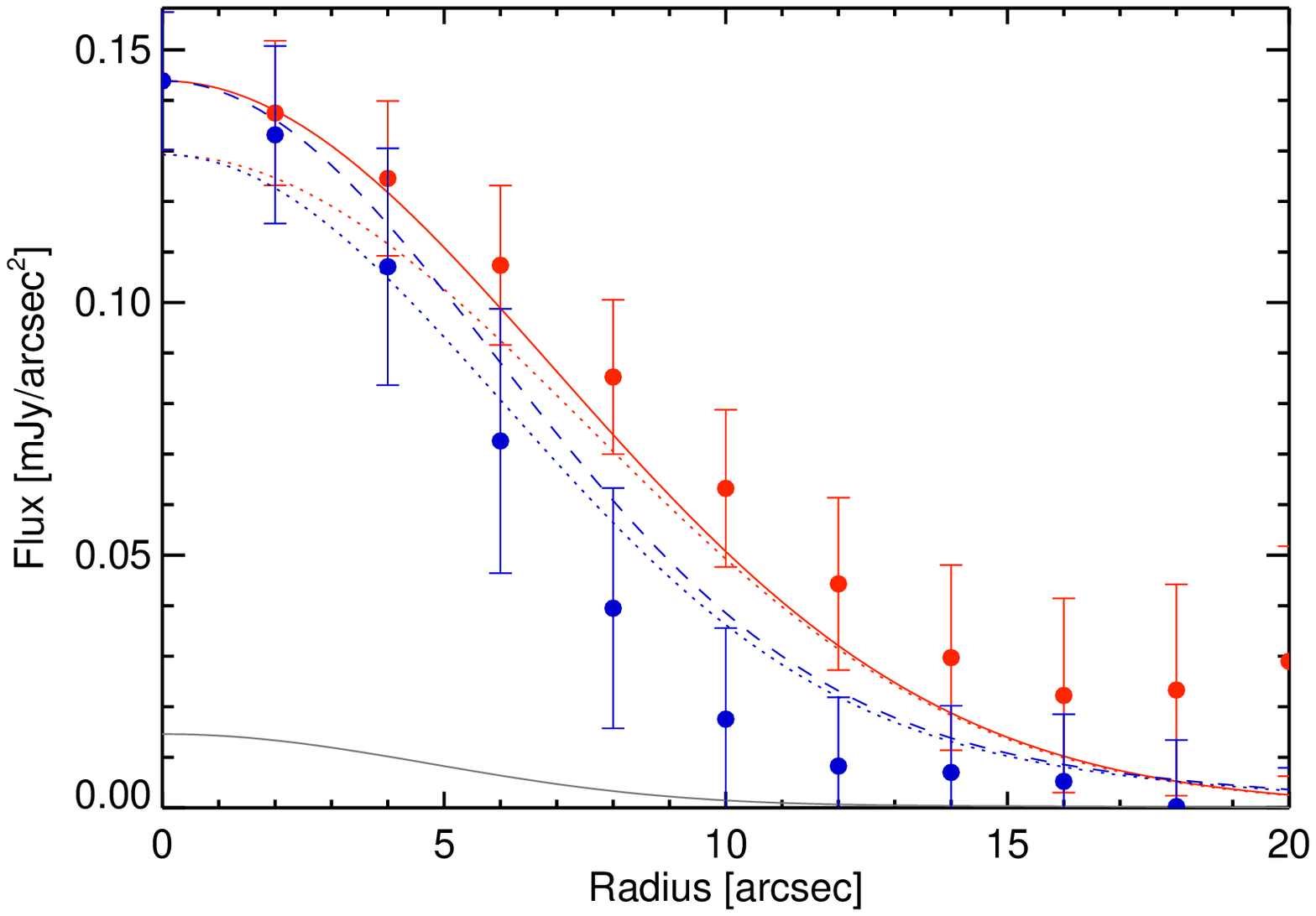}}
\subfigure{\includegraphics[width=0.33\textwidth]{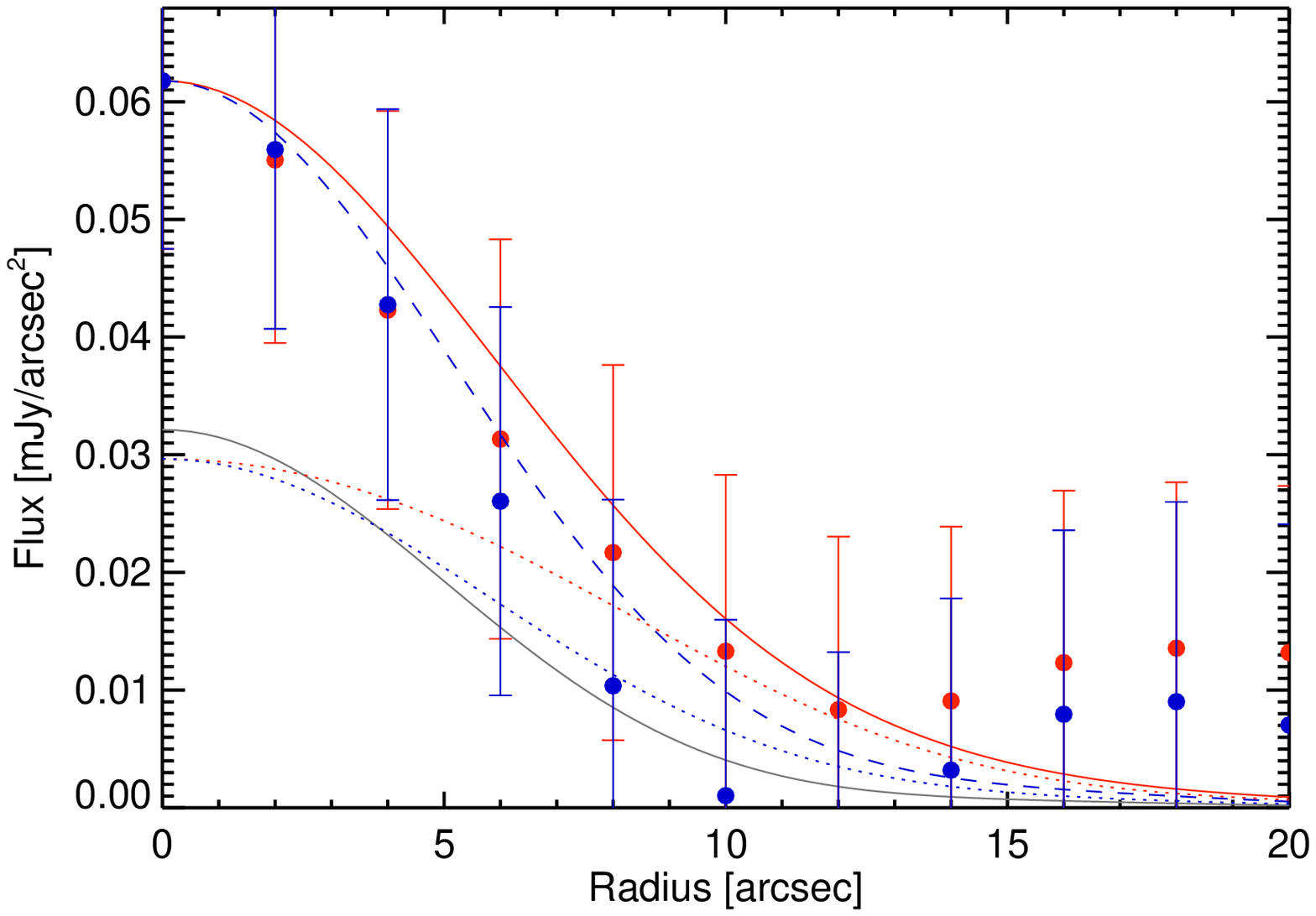}}
\caption{{\sc GRaTeR} results: \textit{Top row}: Spectral energy distributions of HIP~22263, HIP~62207, and HIP~72848 (left to right), calculated from the \textit{simultaneous} best fit model. Empty data points are photosphere measurements, black data points are ancillary mid- and far-infrared measurements, green data points are the flux densities used to scale the photosphere model, dark blue data points are flux densities extracted from the \textit{Spitzer} IRS spectrum, represented by the light blue data points, and red data points are the \textit{Herschel} PACS and SPIRE flux densities. The combined photosphere and disc model is a solid black line. The disc and photosphere model components are represented by dotted and dashed lines, respectively. \textit{Lower rows}: Observed radial profiles of each target at 70, 100, and 160~$\mu$m (top to bottom). The major and minor axis measured values are denoted by the red and blue data points, respectively. The model major and minor axis radial profiles are represented by the red solid line and blue dashed line. The disc contribution is represented by the dotted lines and the stellar contribution is shown as a solid grey line.}
\label{grater_plots}
\end{figure*}

In this section we summarize the results of the modelling processes described in the previous section on a target by target basis. The measured position angles and inclinations of the three discs are consistent across the different methods and therefore require no further comment. The model disc images derived from the modified blackbody analysis are presented in Fig. \ref{mbb_plots} whilst the disc SEDs and radial profiles of the {\sc GRaTeR} analysis are presented in Fig. \ref{grater_plots}. The modified blackbody and {\sc GRaTeR} analyses are broadly in agreement and the results presented in Fig. \ref{mbb_plots} are therefore illustrative of the outcome from both methods. A summary of the best fit model parameters for all three methods is presented in Table \ref{mod_table}.

\begin{table}[ht!!]
\caption{Modelling results summary; see text for details of parameters fitted in each method.}
\centering
\begin{tabular}{lrrr}
\hline\hline
Parameter & HIP~22263 & HIP~62207 & HIP~72848 \\
\hline
 & \multicolumn{3}{c}{Single annulus}\\
\hline
$R_{\rm in}$ [au] & 44~$\pm$~1 &  \phantom{0}82~$^{+2}_{-3}$ & 63~$^{+1}_{-6}$ \\
$\theta$ [\degr] & 10~$^{+3}_{-9}$ & 106$^{+5}_{-2}$ & 58~$^{+4}_{-1}$ \\
$i$ [\degr] & 50~$\pm$~2 & \phantom{0}63~$\pm$~2 & 85~$^{+5}_{-4}$ \\
$\chi^{2}_{\rm red}$ & 1.16 & 1.45 & 1.29 \\
\hline
 & \multicolumn{3}{c}{Modified blackbody$^{a}$}\\
\hline
$R_{\rm in}$ [au] & 22 & 72 & 74 \\
$R_{\rm out}$ [au] & 116 & 82 & 84 \\
$\tau_0~[\times 10^{-5}]$ & 8.2 & 38 & 11 \\
$\alpha$ & -0.6 & 0 & 0 \\
$T_{\rm 1au}$ [K] & 416 & 515 & 515 \\
$f_T$ & 1.5 & 1.8 & 2.2 \\
$\lambda_0$ [$\mu$m] & 57 & \ldots & \ldots \\
$\beta$ & 0.7 & \ldots & \ldots \\
$\theta~[^\circ]$ & 7 & 105 & 64 \\
$i~[^\circ]$ & 51 & 56 & $>$70 \\
$L_{\rm IR}/L_\star~[\times 10^{-5}]$ & 3.1 & 2.4 & 1.0 \\
$\chi^2_{\rm red}$ & 1.40 & 1.20 & 2.40 \\
\hline
 & \multicolumn{3}{c}{{\sc GRaTeR}$^{b}$}\\
\hline
$r_{0}$ [au] & 20.9~$\pm$~4.5 & 53.7$^{+9.7}_{-14.7}$ & 57.8~$\pm$~20.0\\
\ldots & [20.8] & [53.7] & [66.9] \\
$s_{\rm min}$ [$\mu$m] & 1.89$^{+0.48}_{-0.43}$ & 5.17$^{+1.17}_{-0.56}$ & 1.30$^{+2.60}_{-1.25}$ \\
\ldots & [2.15] & [5.86] & [0.12] \\
$\alpha$ & -1.00$^{+0.61}_{-0.58}$ & -0.50$^{+0.25}_{-0.87}$ & -5.00$^{+2.50}_{-0.00}$ \\
\ldots & [-0.50] & [-1.00] & [-5.00] \\
$\gamma$ & 3.9$^{+0.2}_{-0.1}$ & 4.3$^{+0.3}_{-0.1}$ & 3.7$^{+0.8}_{-0.2}$ \\
\ldots & [3.9] & [4.3] & [3.5] \\
$M_{\rm dust}$ [$\times10^{-4}M_{\oplus}$] & 6.06~$\pm$~0.75 & 32.2$^{+5.8}_{-6.5}$ & 1.48$^{+1.74}_{-0.82}$ \\
\ldots  & [6.10] & [26.9] & [2.00] \\
$L_{\rm IR}/L_{\star}$ [$\times10^{-5}$] & 2.9 & 2.1 & 0.3 \\
$\chi^{2}_{\rm red}$ & 0.65 & 0.59 & 0.33 \\
\hline
\end{tabular}
\tablefoot{(a) The $\chi^{2}_{\rm red}$ for HIP~72848 is higher because the background wasn't added to the model, and there are nearby sources with significant emission. The modified part of the blackbody is not needed for HIP~62207 and HIP~72848, denoted by `\ldots' for these parameter values. The ring width was fixed to 10~au for HIP~62207 and HIP~72848.\\
(b) The number in square brackets denotes the best fit value for each parameter when fitting all parameters simultaneously.}
\label{mod_table}
\end{table}

\subsection{HIP~22263}

For HIP~22263, the single annulus fitting of this disc suggests a radius of 44~au, whilst the modified blackbody modelling finds a best fit solution of a disc extending from 22 to 116~au. A reasonably flat surface density profile with $\alpha\!=\!-0.6$ is suggested by the black body model but this is not significant due to the marginally resolved disc providing only a minimum of information on its spatial extent. However, we can preclude the presence of large amounts of dust at great distances from the star which would produce a more extended source emission and act to flatten the disc surface density from longer wavelength observations. From the image analysis summarized in Table \ref{orient}, we see that the disc is marginally resolved at 70 and 100~$\mu$m, presenting a symmetric morphology aligned approximately north-south. The lack of extended emission at 160~$\mu$m also suggests that the disc is relatively compact and lacks a very cold extended component, which has implications for the shape of the SED.

The best fit {\sc GRaTeR} model for HIP~22263 is a relatively small disc; the simultaneous best fit model to the SED and radial profiles has a $\chi^{2}_{\rm red}$ of 0.65 with 67 degrees of freedom (see Fig. \ref{grater_plots}, left hand column). The disc peak surface density lies at $r_{0}~=~20.8~$au, smaller than the single annulus model but consistent with the inner radius of the modified black body model. This is likely the result of considering the SED in the fitting process of the modified black body and {\sc GRaTeR} methods. The surface density exponent, $\alpha\!=\!-0.5$, suggests an outwardly decreasing surface density profile, also in line with the modified black body analysis. We obtain a minimum grain size, $s_{\rm min}\!=\!2.15~\mu$m, a few times larger than the blow-out radius (0.5~$\mu$m). The particle size distribution exponent, $\gamma\!=\!3.9$, is comparable to that of a system in steady state collisional cascade \citep[3.5 to 3.7,][]{dohnanyi69,ta07,gaspar12} and within the range of expected values for typical debris discs (3 to 4) \citep[see e.g.][]{panschl12}. The implied dust mass, $M_{\rm dust}\!=\!6.1\times10^{-4}M_{\oplus}$, and disc fractional luminosity, $2.9\times10^{-5}$, puts this disc at the brighter end of those observed by DUNES with comparable brightness to e.g. HIP~17439 \citep{ertel14} and HIP~32480 (Stapelfeldt et al., in prep.), but is not exceptional. The disc orientation has a position angle $\theta\!=\!4.7~\pm~2.1\degr$ and inclination $i\!=\!51~\pm~10\degr$, which are well constrained by the Gaussian fit to the 100~$\mu$m image.

Although the proposed disc model replicates the mid- and far-infrared photometry well, it begins to diverge beyond 250~$\mu$m, as the measured flux densities become brighter than the predicted model flux densities at both 350 and 850~$\mu$m. This could be due to a number of reasons, the two most likely of which are contamination by a background object, or the presence of a second, very cold, dust component to the disc. Firstly, a halo of very small grains would be unusual around a mature, solar type star. Furthermore, to fit the SED, the halo would need a temperature of $\leq~20~$K, which would place it at a distance of (at least) $\sim$~200~au, beyond the largest discs identified around similar G type stars, e.g. HD~107146 \citep{ertel11} or HD~207129 \citep{loehne12}, and producing extended emission at wavelengths up to 250~$\mu$m, which is not observed. These properties would place it amongst those 'cold disc candidates' identified in \cite{eiroa11}, \cite{eiroa13}, and \cite{marshall13}, albeit much brighter than those previously identified and therefore inconsistent with the explanation for the phenomenon proposed in \cite{krivov13}. We therefore disfavour this explanation for the shape of the SED. In the case of contamination, a temperature of 20~K is typical of background galaxies at redshifts from $z\!=\!1$--$2$ \citep{magnelli13}. Two red sources are visible to the south of HIP~22263 in the PACS 160~$\mu$m map at separations of 25\arcsec~and 50\arcsec. The second of these has a bright counterpart in both the SPIRE 350 and 500~$\mu$m maps. We therefore infer that the 350~$\mu$m flux density is partially contaminated by the interloper and that, given the number of background sources in proximity to HIP~22263, the JCMT/SCUBA observation may similarly be a victim of contamination or misidentification, leading to the high flux density measurement reported at 850~$\mu$m \citep{greaves09}. High angular resolution observation of HIP~22263 at (sub-)mm wavelengths could confirm this interpretation using e.g. the AzTEC instrument on the Large Millimetre Telescope ($\lambda\!=\!1.1~$mm, FWHM = 8\arcsec, $\sigma_{\rm rms}\!=\!0.2~$mJy).

\subsection{HIP~62207}

The single annulus and modified blackbody models of HIP~62207 suggest a disc radius of 82 or 72~au, respectively. In both cases a narrow annulus with a width of $\leq$~10~au is the preferred extent, albeit by design in the case of the single annulus model. The PSF subtracted images exhibit significant extended emission centred on the star, but also highlights the presence of a contaminating background source to the northwest of the disc, particularly at 160~$\mu$m (see Figs. \ref{disc_imgs} and \ref{mbb_plots}, middle column). This background source was modelled as a scaled PSF and subtracted from the observations before modelling the disc to avoid contaminating the radial profile measurements.

Conversely, we find with {\sc GRaTeR} that the best fit disc model for HIP~62207 is a cool, broad ring; the simultaneous best fit model to the SED and radial profiles has a $\chi^{2}_{\rm red}$ of 0.59 with 40 degrees of freedom (see Fig. \ref{grater_plots}, middle column). The disc has a peak surface brightness at $r_{0}\!=\!53.7$~au and an outwardly decreasing surface density, i.e. $\alpha\!=\!-1.00$. This mismatch between the previous approaches, which are in good agreement, and the {\sc GRaTeR} results is likely due to the dust grain temperature and radial location being self consistent in the {\sc GRaTeR} modelling through the assumption of the dust grain properties, a detail which is neglected by the two simpler approaches. The disc around HIP~62207 is more elongated than either of the other discs considered here, so the surface density measurement is meaningful for interpretation of the dust distribution within the disc. We obtain a minimum grain size, $s_{\rm min}\!=\!5.86~\mu$m, around ten times larger than the blow-out radius ($s_{\rm blow}~\sim$ 0.5~$\mu$m) and comparable to the size ratio calculated for the disc around HD~207129 \citep{loehne12} rather than the more typical debris discs where the ratio of $s_{\rm min}/s_{\rm blow}$ lies closer to values of around two to five \citep[][and Pawellek et al., submitted]{krivov06,ta07,tw08}. The particle size distribution exponent value of $\gamma\!=\!4.3$ is somewhat steeper than the range of expected values for typical debris discs (3 to 4), and much steeper than classical values for a steady state collisional cascade (3.5 to 3.7). This is not to imply we are observing a system that is out of collisional equilibrium although its value of $\gamma$ does approach that of HIP~114948 ($\gamma~=~4.7$), one of the 'steep SED' discs identified in \cite{ertel12}. The dust mass implied from the SED fit, $2.69~\times10^{-3}~M_{\oplus}$, and dust fractional luminosity, $2.1\times10^{-5}$, are comparable to other discs in the DUNES sample. Caution should be used in comparing the dust mass of HIP~62207 with those of other works as the steeper size distribution and large minimum grain size might be responsible for the large mass value derived here. We note that the size distribution measured for dust grains will not apply to the larger bodies which do in fact dominate the total mass of the disc. The disc orientation, with a position angle $\theta\!=\!111.2~\pm~1.3\degr$ and inclination $i\!=\!56~\pm~10\degr$, is well constrained from the 2D Gaussian profile fits to the extended emission in the PACS images. 

The best fit model of HIP~62207, with its large minimum grain size, broad disc and flat surface density, has similarities to that of HD~207129's disc \citep{loehne12}. Likewise, the sub-solar stellar metallicity in combination with the presence of an extended debris belt and lack of known Jovian mass companion(s) also parallels the HD~207129 and $\tau$ Ceti systems, making this star a good candidate system to search for low mass exoplanets \citep{marshall14}. The lack of sub-mm measurements for the target limits the detail which can be drawn from the modelling process, whilst further imaging observations at high(er) angular resolution are key to pinning down the disc morphology.

\subsection{HIP~72848}

In the case of HIP~72848, the single annulus model suggests a radial extent of 63~au for the disc when combining information from all three PACS bands. If the 70~$\mu$m image is omitted from the fitting, the preferred radius increases to 74~$^{+4}_{-6}$~au, in good agreement with the modified black body radius. Due to the marginal extension of this disc interpretation of the surface density profile is poorly constrained from the available data, as was the case for HIP~22263. The images of HIP~72848 therefore exemplify the difficulty in the interpretation of this disc as the extended emission has a different orientation at 70 and 100~$\mu$m (see Fig. \ref{disc_imgs} and Table \ref{orient}). This is a result of the disc being faint and its emission at 70~$\mu$m being dominated by the large contribution of the star to the total flux density and therefore the shape of the source brightness profile, which is not a simple Gaussian. 

We find that the simultaneous best fit disc model from {\sc GRaTeR} for HIP~72848 is a cold, narrow disc with a $\chi^{2}_{\rm red}\!=\!0.33$ with 35 degrees of freedom (see Fig. \ref{grater_plots}, right hand column). The individual best fit peak in disc surface brightness is at $r_{0}\!=\!66.9$~au, consistent with the estimates from the other modelling approaches. The radial fall off for the surface density $\alpha\!=\!-5.0$ is very steep for both the simultaneous and individual best fit values, {suggestive of a narrow disc. However the disc is only marginally resolved such that, similar to HIP~22263, the meaningfulness of this result is weak in terms of constraining the disc architecture. We obtain a minimum grain size, $s_{\rm min}\!=\!0.12~\mu$m. The small minimum grain size for this disc may reflect that those grains are not being removed from the system via radiation pressure blow-out, which would not be expected to be an effective removal mechanism for dust grains around a K star due to their low luminosity \citep{kw13}. An extended halo of small grains on highly eccentric bound orbits would act to blur the outer edge of the disc, broadening its observed radial profile. We can calculate the eccentricity of grains with sizes of 0.12 and 1.3~$\mu$m and assumed optical properties using the $\beta$ parameter which results in orbital eccentricities of 0.25 and 0.15, respectively. The grains are not therefore expected to have large radial excursions from their radius of production, which runs counter to the possible presence of an extended halo. The particle size distribution exponent $\gamma\!=\!3.5$ is consistent with a steady state collisional cascade. The dust mass implied from the simultaneous best fit model to the data is 2.0~$\times10^{-4}$~$M_{\oplus}$. The dust fractional luminosity of 2.8$\times10^{-6}$ is at the fainter end of discs in the DUNES survey which, combined with the disc's warmer blackbody temperature, explains the difficulty in obtaining accurate modelling results for this disc. The disc position angle $\theta\!=\!66.4~\pm~2.0\degr$ is derived from the 100~$\mu$m surface brightness profile, as is the inclination, $i~\ge~84~\pm~10~\degr$, which is a lower limit due to the lack of extended emission along the source minor axis.

The dominant contribution to the flux density at 70~$\mu$m is the star, which makes interpretation of the extended emission tricky at this wavelength as any asymmetries and extended emission are more strongly tied to the stellar PSF than the dust. Further high S/N imaging of the disc would therefore help clarify the orientation and extent of the disc, which are inconsistent between the 70 and 100~$\mu$m images. Additionally, the SED is poorly sampled as it rises from the stellar photosphere, which leaves the modelling sensitive to the choice of inclusion of the \textit{WISE} data point at 22~$\mu$m in our fitting. The inclusion of this data point in the modelling favours a larger minimum grain size of 0.42~$\mu$m (2.1~$\mu$m for the individual fit) but the disc structure remains the same -- this difference between the simultaneous and individual best fit values of $s_{\rm min}$ is symptomatic of the deficiencies in the data set available for modelling this disc. In this regard, a greater density of photometric points covering the mid-infrared to trace the rise of the disc SED above the stellar continuum would be useful, moreso perhaps than sub-mm photometry, which this star lacks.

\begin{figure*}[ht!!]
\subfigure{\includegraphics[width=0.32\textwidth]{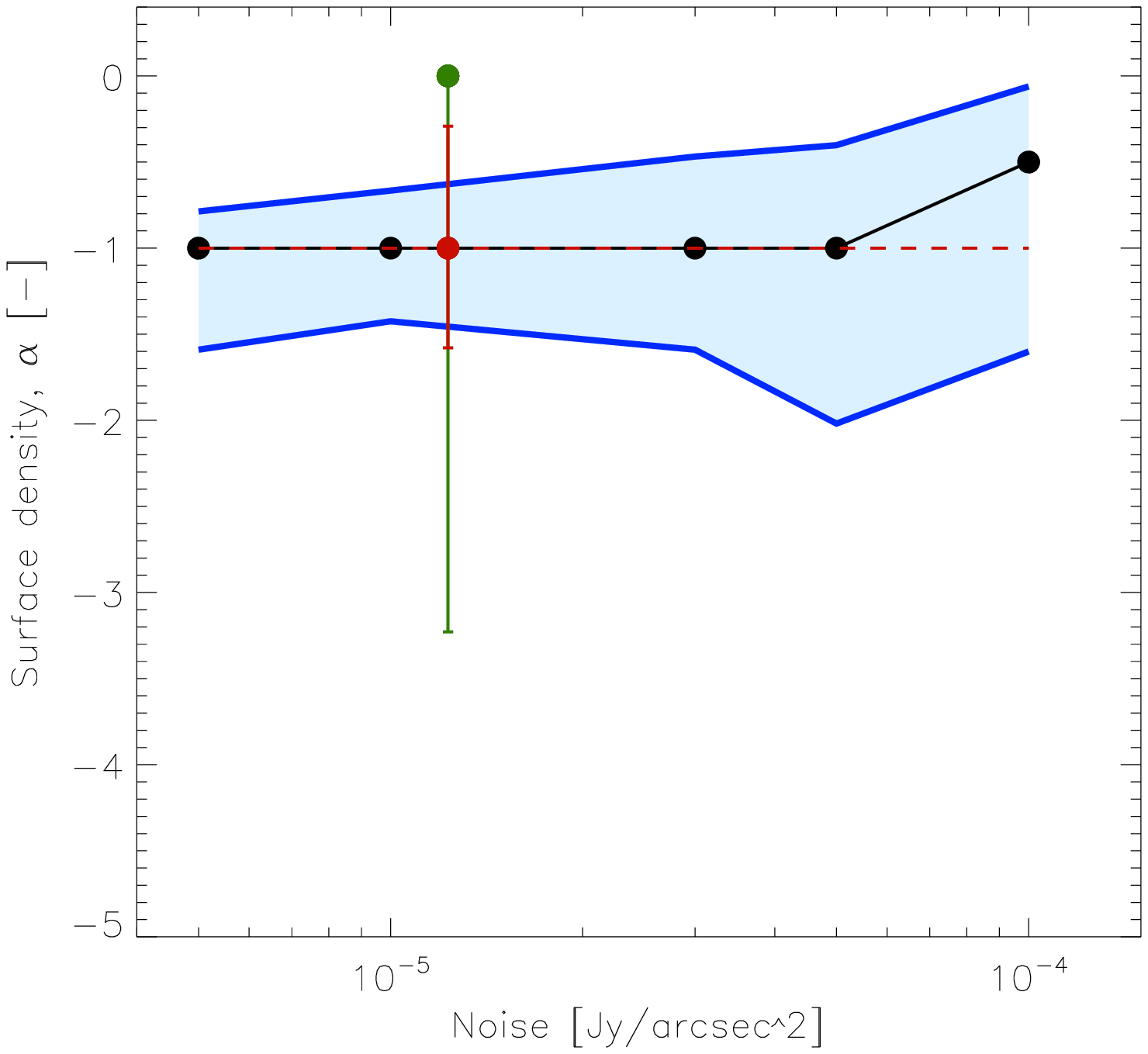}}
\subfigure{\includegraphics[width=0.32\textwidth]{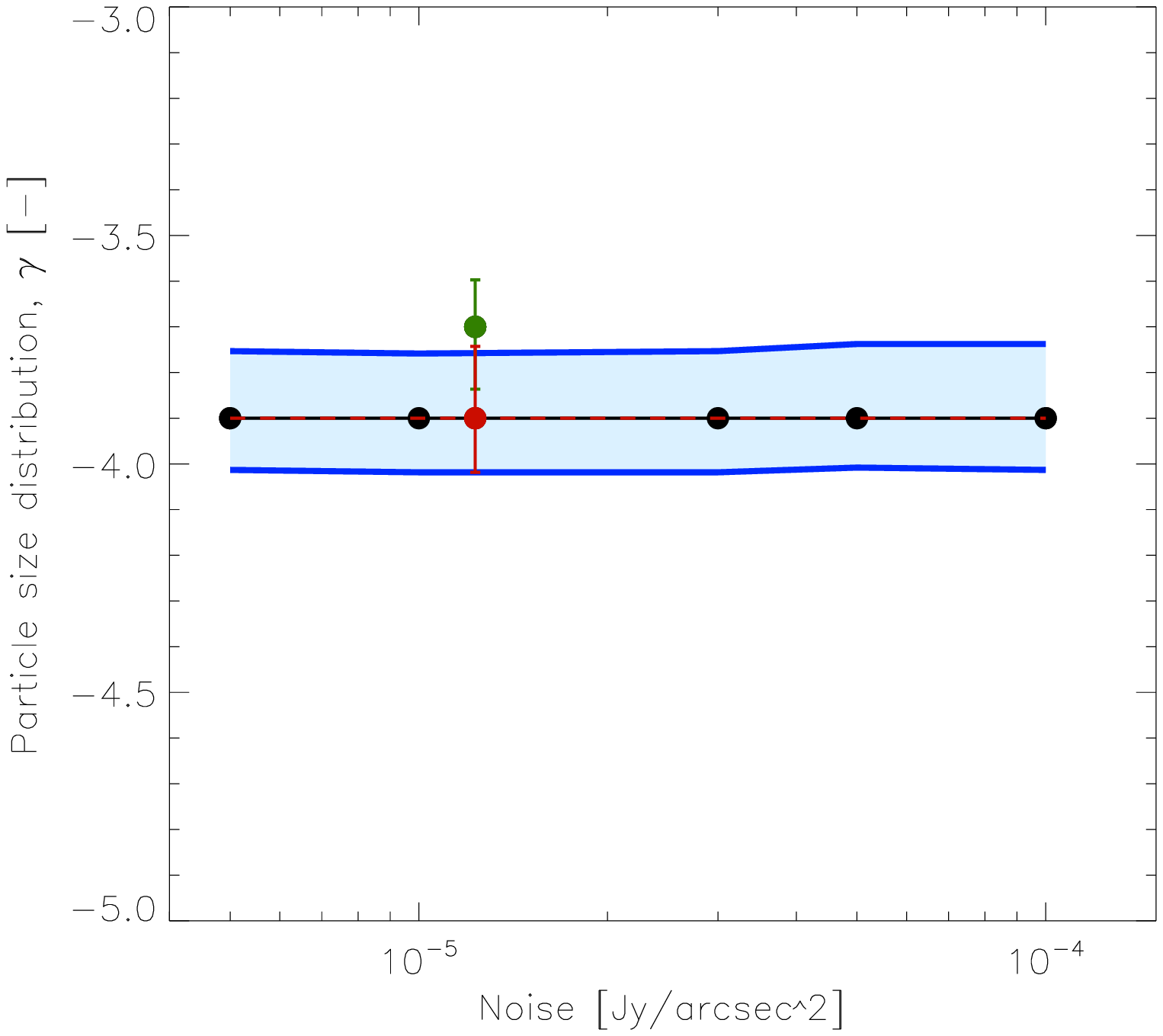}}
\subfigure{\includegraphics[width=0.32\textwidth]{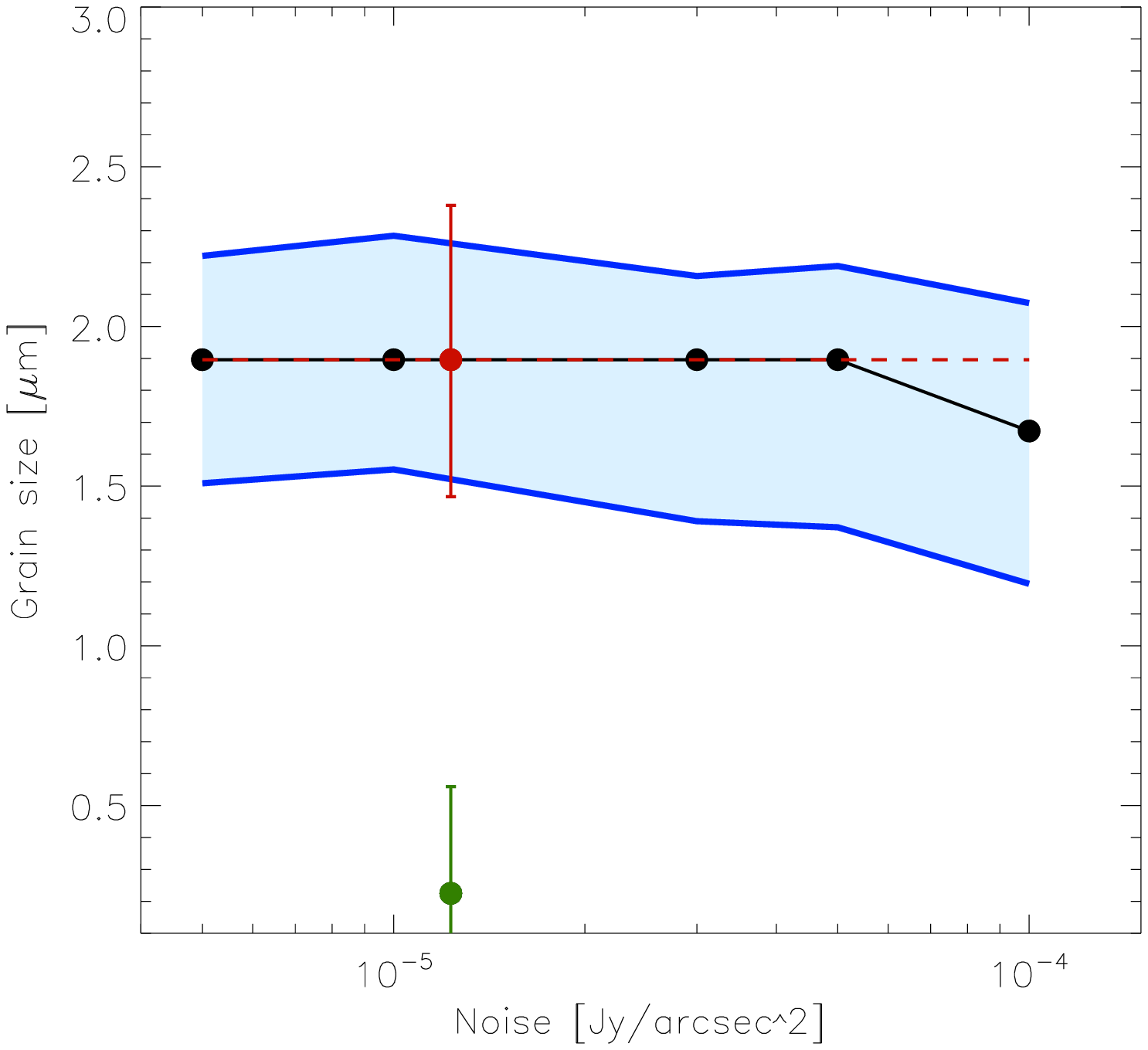}}\\
\subfigure{\includegraphics[width=0.32\textwidth]{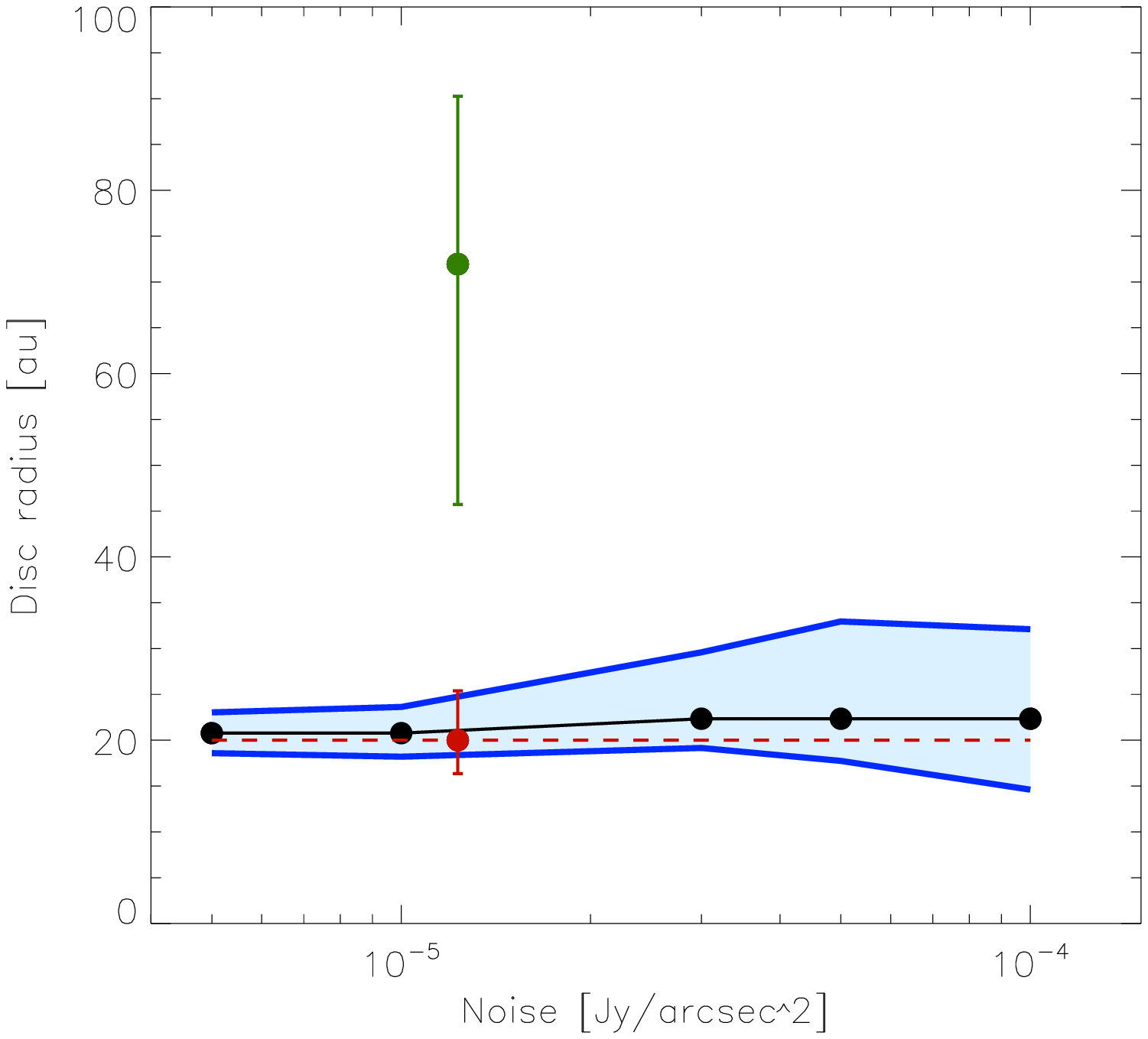}}
\subfigure{\includegraphics[width=0.32\textwidth]{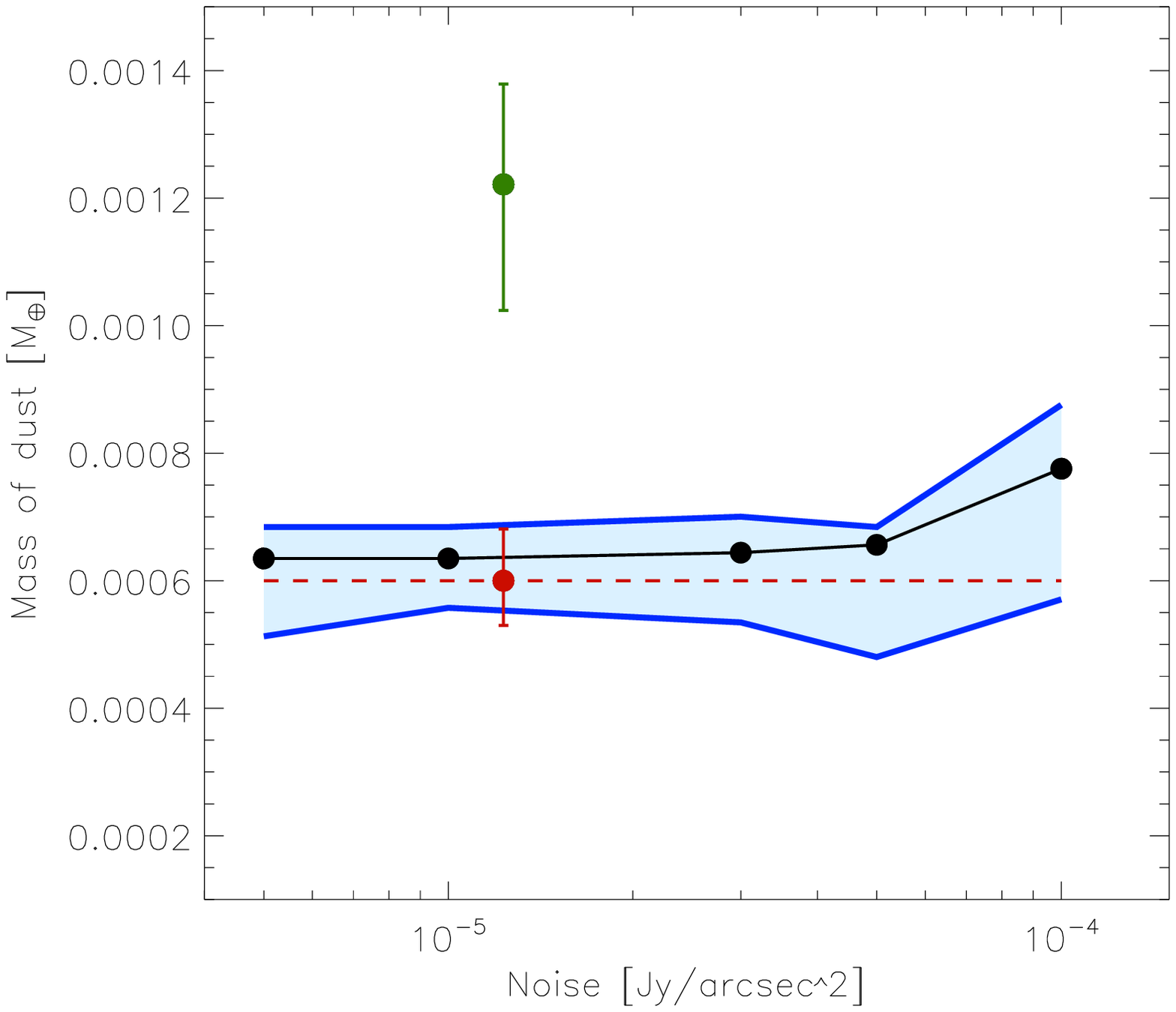}}
\caption{Recovered model parameter values as a function of image noise for the first model, based on the best fit to observations of HIP~22263.  \textit{Top row}: Surface density, $\alpha$, particle size distribution, $\gamma$, and minimum grain size, $s_{\rm min}$. \textit{Bottom row}: Disc peak radius, $r_{\rm 0}$, and dust mass, $M_{\rm dust}$. The red data point denotes the results for the observed SED and radial profiles, with the red dashed line denoting the input value to the model for each parameter. The green data point denotes a fit to only the observed SED. The black data points denote models fitted to both the SED and radial profiles with increasing noise contributions with the envelope of uncertainties for the models marked by the light blue shaded region, bounded by the dark blue lines. \label{results_model1}}
\subfigure{\includegraphics[width=0.32\textwidth]{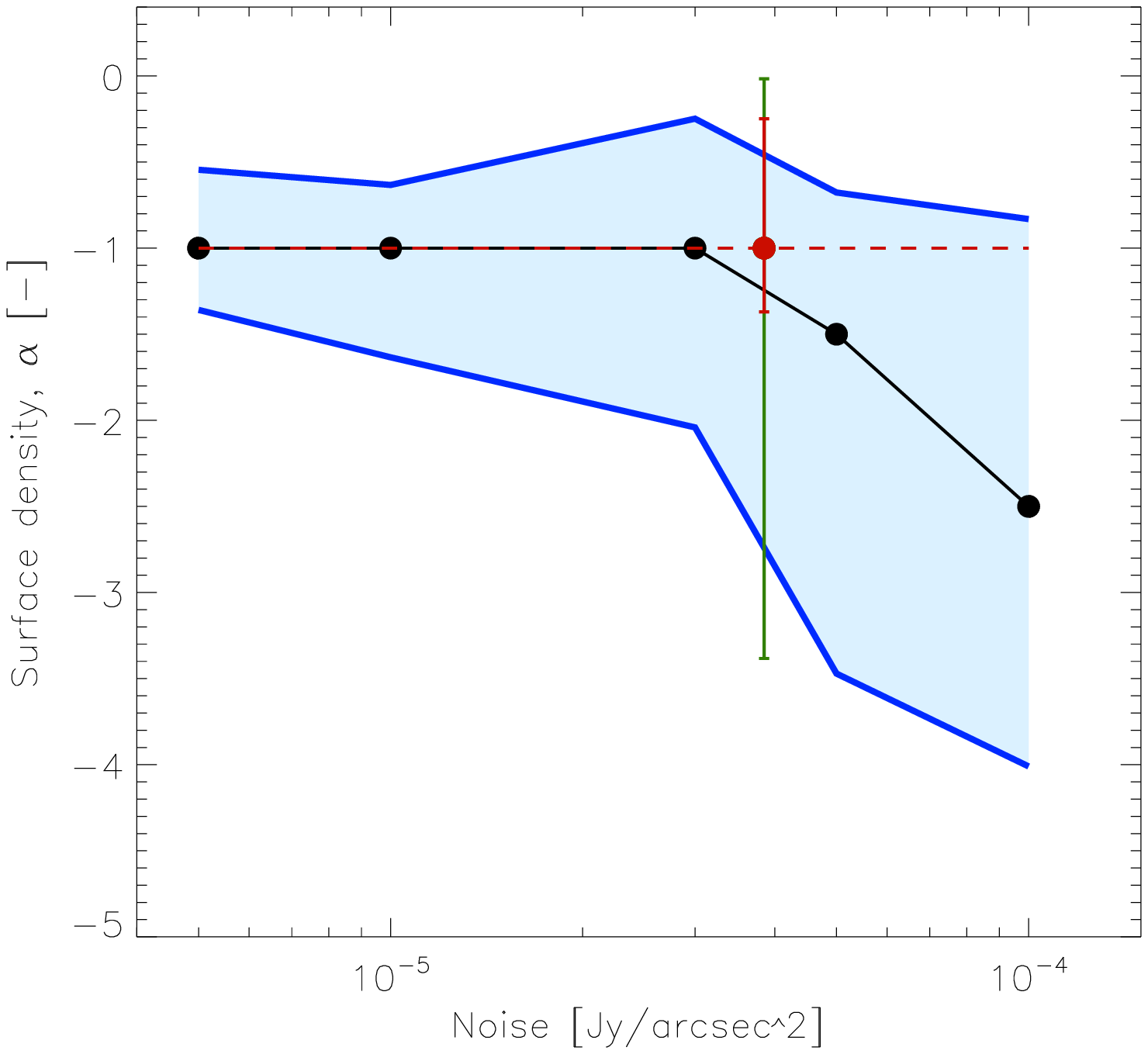}}
\subfigure{\includegraphics[width=0.32\textwidth]{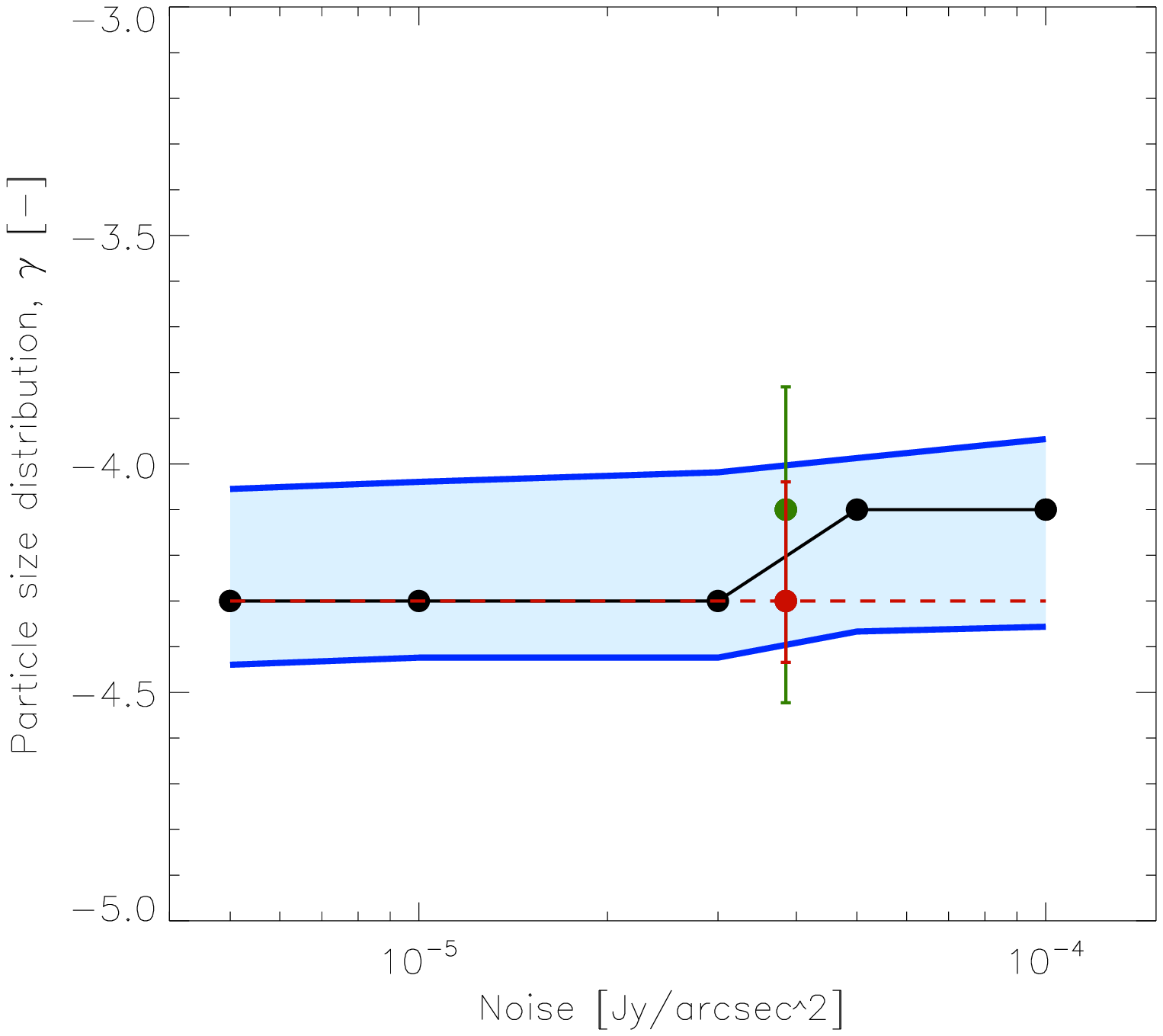}}
\subfigure{\includegraphics[width=0.32\textwidth]{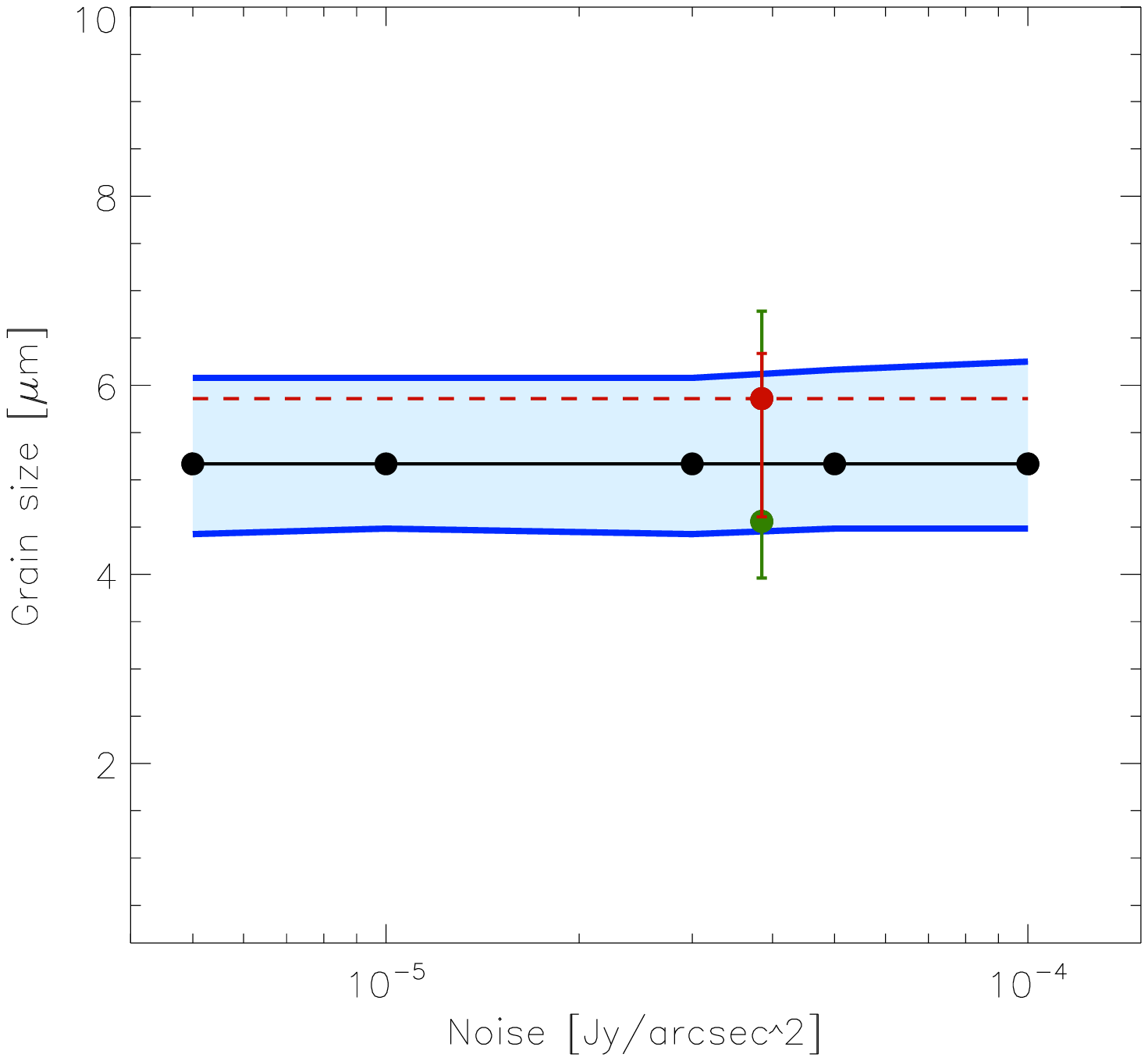}}\\
\subfigure{\includegraphics[width=0.32\textwidth]{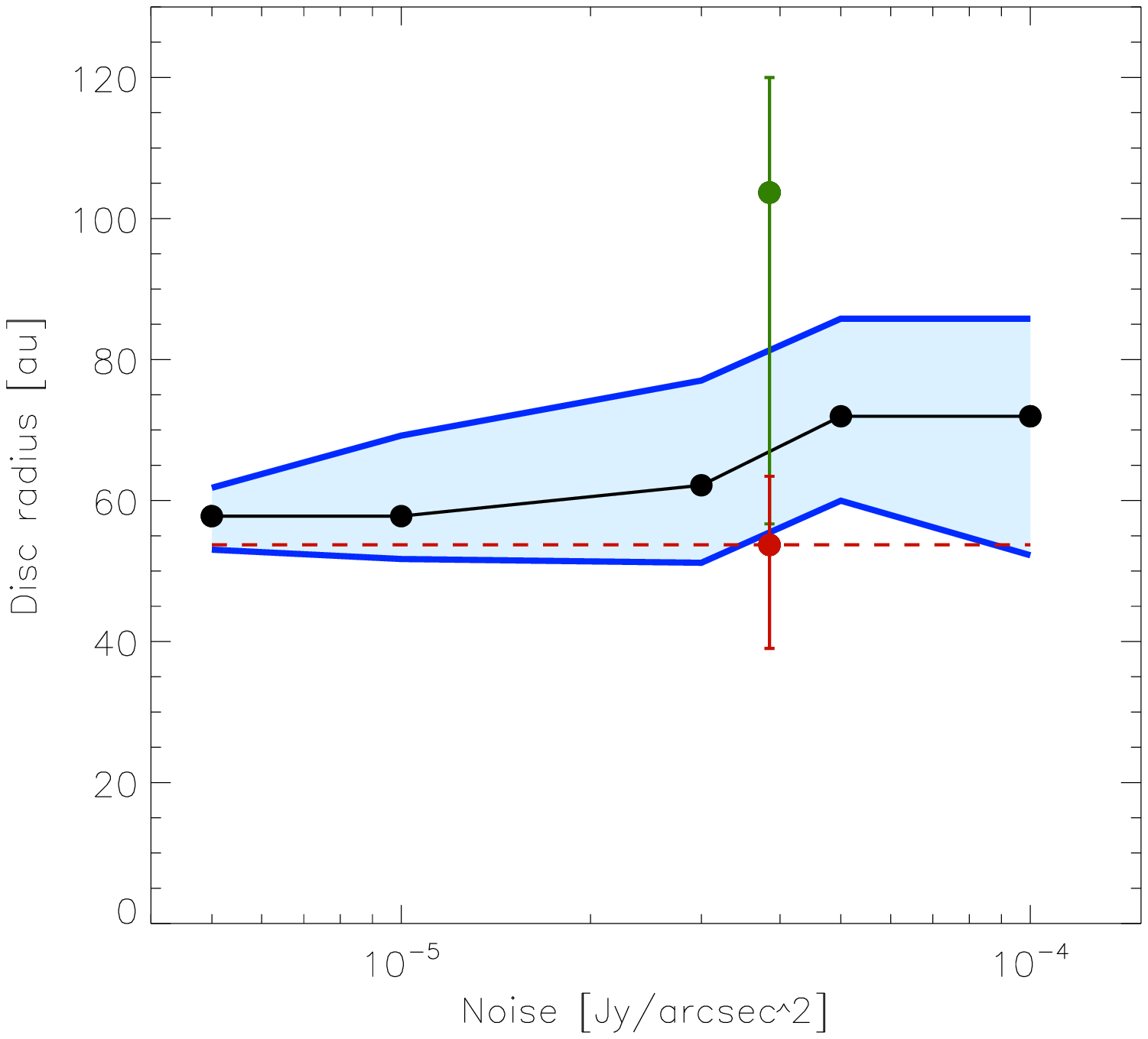}}
\subfigure{\includegraphics[width=0.32\textwidth]{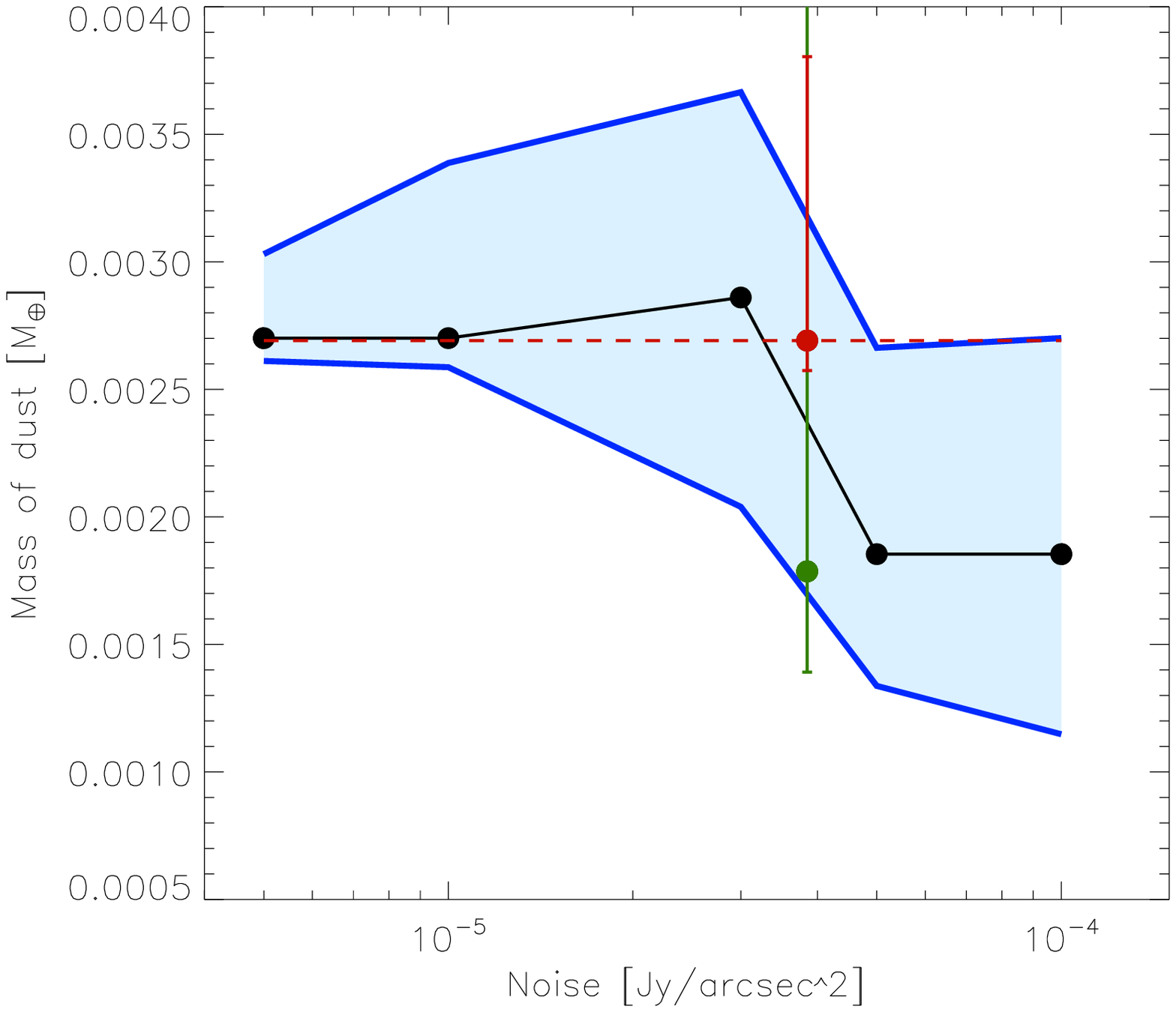}}
\caption{Recovered model parameter values as a function of image noise for the second model, based on the best fit to observations of HIP~62207. \textit{Top row}: Surface density, $\alpha$, particle size distribution, $\gamma$, and minimum grain size, $s_{\rm min}$. \textit{Bottom row}: Disc peak radius, $r_{\rm 0}$, and dust mass, $M_{\rm dust}$. See above for an explanation of the figure. \label{results_model2}}

\end{figure*}

\section{Interpreting extended emission}  \label{sect_int}

In this section we seek to quantify the influence that knowledge of the disc radial profiles has on how accurately the dust parameters of the disc can be determined. To this end we consider two cases of extended emission at moderate S/N: that of HIP~22263 (hereafter Model 1), representative of a marginally extended disc along its major axis, and HIP~62207 (hereafter Model 2), representative of a disc clearly extended along its major axis. These two cases are representative of the majority of extended debris discs observed by \textit{Herschel}, see e.g. \cite{booth13}, \cite{eiroa13} and Pawellek et al., submitted.

We begin by taking the best fit models of HIP~22263 and HIP~62207 as determined from the {\sc GRaTeR} results as our base reference for the two models, creating an image for the dust distribution and emission for each PACS band. To the model images we add the appropriate stellar photosphere contribution for each wavelength before convolution with an instrument PSF. We then add normally distributed random noise component to the convolved images, with standard deviations bracketing values measured in \textit{Herschel} PACS images in five increments spanning 5$\times10^{-6}$, 1$\times10^{-5}$, 3$\times10^{-5}$, 5$\times10^{-5}$, and 1$\times10^{-4}$ Jy/arcsec$^{2}$. The source radial brightness profiles for each set of noisy images are then measured as per the observations (see Section 2.4). These simulated profiles are then input into {\sc GRaTeR} along with the observed SED, since we only want to trace the influence of the extended emission, and the fitting procedure is run again, as described for the observations (see Section 3). For each model we also produce a result using only the SED as a constraint in the modelling process. The disc parameters obtained from fitting both sets of noisy models (and the SED only ones) are subsequently compared to the input values.

From Figs. \ref{results_model1} and \ref{results_model2} we see the model input parameters were better recovered by the fitting process for the images with lower noise, which is exactly what one would expect. We note several trends in the match between the model input parameters and the values recovered by the fitting procedure which hold despite the difference in the two cases considered here. 

For the Model 1 set the radial surface density exponent, $\alpha$, was broadly flat over the range of noise values considered here and replicated the input model well. By contrast, the best fit value of the Model 2 set tended towards lower $\alpha$ values as image noise increased. The additional constraint of the sub-millimetre data in the source SED of Model 1 may be responsible for this difference in outcome.

The disc inner radius, $r_{0}$, tended to be overestimated compared to the input model with the degree over of estimation decreasing for smaller noise values. Model 1, which had sub-millimetre photometry, had smaller error bars by a factor of two compared to Model 2 on the fitted values for this parameter. In either case, if no radial profiles were used to constrain the model fitting, $r_{0}$ was overestimated by a factor of 2--4. 

The minimum grain size, $s_{\rm min}$, was well recovered for all cases which used radial profiles to constrain the fitting, independent of noise. In the SED-only fits the value of $s_{\rm min}$ is underestimated. The interplay between the estimation of $r_{0}$ and $s_{\rm min}$ whose properties are degenerate in the model fitting is well known and serves to emphasize the importance of spatially resolved imaging in accurate determination of disc properties. 

The quality of the model fit for the particle size distribution exponent, $\gamma$, was good, independent of the noise which is quite expected as the determination of this parameter is sensitive to the SED rather than the radial profiles.

The dust mass was somewhat variable. The SED-only fitting resulted in a higher mass than the model fits using radial profiles for both Model 1 and 2. The case of Model 1 with sub-mm photometry resulted in better recovery of the input parameters. The dominant effect on the dust mass for both Model 1 and 2 was the surface density, as can be seen in the correlation between the variation of $\alpha$ and $M_{\rm dust}$ in Figs. \ref{results_model1} and \ref{results_model2}. 

\section{Conclusions}  \label{sect_con}

We present an analysis of the far-infrared extended emission from the circumstellar debris discs around HIP~22263, HIP~62207, and HIP~72848 by three separate methods. In each case we find that the disc architecture is well represented by a single, cold dust annulus. Simultaneous fitting of the SED and radial profiles provides better constraint of the physical properties of the constituent dust than was heretofore possible. 

Using these new resolved images we identify evidence of potential contamination in the images of two of the three sources. In the case of HIP~22263, the model disc is a good match to the far-infrared profiles, and photometry data up to 250~$\mu$m, but beyond that the disc model under predicts the observed flux density. We believe that this is an instance of background contamination at sub-mm wavelengths; high S/N, high angular resolution observations at millimetre wavelengths would be required to accurately trace the decline of the dust emission from the disc avoiding background contamination. For HIP~62207, the two nearby contaminating sources are easily disentangled from the extended disc emission at far-infrared wavelengths, particularly 100 and 160~$\mu$m. Our modelling reveals the disc to have a near flat surface brightness and large radial extent, spanning $\sim~$50--150~au, comparable to the extent of the disc recently resolved around HIP~17439 \citep{ertel14} and reminiscent of the disc around HD~207129 \citep{marshall11,loehne12}. Finally, our modelling of HIP~72848 suggests a cold, narrow debris disc, although any definitive interpretation is hampered by the dominated by contribution of the stellar component at 70~$\mu$m to the total observed flux density hampering extraction of the disc architecture from the images and a paucity of mid-infrared and sub-mm photometry leading to a weakly constrained SED as inputs to the modelling process. 


\bibliographystyle{aa}
\bibliography{resolved_discs}


\begin{acknowledgements}
The authors thank the anonymous referee for his constructive criticism. This research has made use of the SIMBAD database, operated at CDS, Strasbourg, France. This research has made use of NASA's Astrophysics Data System. JPM, CE and BM are partially supported by Spanish grant AYA 2011-26202. FK thanks the German \emph{Deut\-sche For\-schungs\-ge\-mein\-schaft, DFG\/} project number WO 857/7-1 for financial support. JCA would like to thank the CNES/PNP for financial support. GMK was supported by the European Union through ERC grant number 279973. MB acknowledges support from a FONDECYT Postdoctral Fellowship, project no. 3140479.
\end{acknowledgements}


\end{document}